\documentclass[fleqn,usenatbib]{mnras}
\usepackage[T1]{fontenc}
\usepackage{amssymb}
\usepackage{amsmath}
\usepackage{multirow}
\usepackage{graphicx}
\newcommand\T{\rule{0pt}{2.6ex}}       % Top strut
\newcommand\B{\rule[-1.2ex]{0pt}{0pt}} % Bottom strut

\title[Constraints on $R$-Process from ESS $^{129}$I$/^{247}$Cm]{Constraints on $R$-process Nucleosynthesis from $^{129}$I and $^{247}$Cm in the Early Solar System}

\author[P. Banerjee et al.]{
Projjwal Banerjee,$^{1}$\thanks{E-mail: projjwal.banerjee@gmail.com}
Meng-Ru Wu,$^{2,3}$
and Jeena S K$^{1}$
\\
$^{1}$Department of Physics, Indian Institute of Technology Palakkad, Kerala 678558, India\\
$^{2}$Institute of Physics, Academia Sinica, Taipei, 11529, Taiwan\\
$^{3}$Institute of Astronomy and Astrophysics, Academia Sinica, Taipei, 10617, Taiwan
}

\date{Accepted XXX. Received YYY; in original form ZZZ}

\pubyear{2015}

\begin{document}
\label{firstpage}
\pagerange{\pageref{firstpage}--\pageref{lastpage}}
\maketitle

\begin{abstract}

%\textbf{another version}

GW170817 has confirmed binary neutron star mergers as one of the sites for rapid neutron capture (\textsl{r}) process. 
However, there are large theoretical and experimental uncertainties associated with the resulting nucleosynthesis calculations and additional sites may be needed to explain all the existing observations.
In this regard, abundances of short-lived radioactive isotopes (SLRIs) in the early solar system (ESS),
that are synthesized exclusively by \textsl{r}-process, can provide independent clues regarding the nature of 
\textsl{r}-process events.
In this work, we study the evolution of \textsl{r}-process SLRIs $^{129}$I and $^{247}$Cm as well as the corresponding reference isotopes $^{127}$I and $^{235}$U at the Solar location. 
We consider up to three different sources that have distinct $^{129}$I/$^{247}$Cm production ratios corresponding to the varied \textsl{r}-process conditions  
in different astrophysical scenarios. 
In contrast to the results found by \citet{Cote2021}, we find that $^{129}$I and $^{247}$Cm in the ESS do not come entirely from a single major event but get contributions from at least two more minor contributors.  
This has a dramatic effect on the evolution of the $^{129}$I/$^{247}$Cm ratio,  
such that the measured ESS value in meteorites may
not correspond to that  
of the \textsl{``last''} major \textsl{r}-process event. 
Interestingly, however, we find that the $^{129}$I/$^{247}$Cm ratio, in combination with the observed $^{129}$I/$^{127}$I and $^{247}$Cm/$^{235}$U ratio in the ESS, 
can still provide important constraints on the properties of proposed \textsl{r}-process sources operating in the Milky Way. 

\end{abstract}


\section{Introduction}
The astrophysical site(s) for the synthesis of heavy elements via the rapid neutron capture process (\textsl{r}-process) has been a long standing puzzle. 
The electromagnetic counterpart
following the detection of gravitational waves from the first ever binary neutron star merger (BNSM) event GW170817 have confirmed the presence of heavy elements in the merger ejecta and thus confirming BNSMs as an \textsl{r}-process source \citep{abbott2017a,abbott2017b,Kasen+17,Cowperthwaite+17,Tanaka+2017}. 
Despite this seminal discovery, very little is known about the details of \textsl{r}-process nucleosynthesis including the possibility of additional sources that have varying levels of neutron richness.
The additional sources may be needed to explain the evolution of \textsl{r}-process elements in the Galaxy as well as neighbouring dwarf galaxies. In particular, the origin of \textsl{r}-process in the early Galaxy as observed in very metal-poor stars (see e.g., a recent review by~\citet{Cowan2019} and the references therein) as well the evolution \textsl{r}-process elements in the disk at late times may require additional sources~\citep{cote2019a}. 
A number of sites have been proposed as candidates for additional \textsl{r}-process sources that include NS -- black hole (BH) mergers~\citep{Lattimer1974,Rosswog2005,Kyutoku2013,Foucart2014}, neutrino-driven winds from core-collapse supernovae (CCSNe)~\citep{Woosley1994,Takahashi1994,Qian1996,Arcones2013}, magneto-rotational supernovae (MRSNe)~\citep{Nishimura2006,Winteler+12,Mosta+17}, collapsars~\citep{Siegel+18,Miller2020}, accretion disk outflows during the common envelope phase of NS--massive star system \citep{Aldana2019}, CCSNe triggered by the hadron-quark phase transitions~\citep{FischerWu2020}, etc.

In this regard, measurement of the abundances of short-lived radioactive isotopes (SLRIs) that are exclusively produced by \textsl{r}-process in the early solar system (ESS), as determined from meteorites as well as Earth's deep-sea sediments, can be used to infer the properties of \textsl{r}-process events that occurred in the Solar neighbourhood~\citep{Wallner_2015,Hotokezaka:2015zea,Bartos:2019cec,Cote2021,Wallner:2021}.
In a recent study \citet{Cote2021} showed %it was shown 
that the observed ratio of the abundances of \textsl{r}-process SLRIs  $^{129}$I and $^{247}$Cm present in the ESS 
can be used to constrain the ``last'' \textsl{r}-process event that contributed to the solar abundance before the formation of the SS. 
In particular, the authors showed that due to the almost identical lifetimes of SLRIs $^{129}$I and $^{247}$Cm, the value of their ratio is insensitive to the uncertainties associated with Galactic chemical evolution and the observed ratio of $438\pm 184$ corresponds to the value produced by the ``last'' \textsl{r}-process event. Additionally, using the fact that  $^{129}$I/$^{247}$Cm is extremely sensitive to the neutron richness, they concluded that the observed value is consistent with medium neutron rich conditions that are most likely associated with the disk ejecta following BNSMs~\citep{Fernandez2013,Just:2014fka,Fujibayashi2018,Miller2019}.
Tidally-disrupted ejecta from NS--BH mergers or BNSMs~\citep{Freiburghaus1999,Goriely2011,Korobkin2012} produce $^{129}$I/$^{247}$Cm ratios that are too low whereas those from the MRSNe result in values that are too high to be compatible with the measurement\footnote{We note that recent BNSM simulations that include the effect of weak interactions generally predict less neutron-rich conditions in the early-time ejecta, particularly for the component that originates from the collisional interface of the two NSs~\citep{Wanajo:2014wha,Radice2016,Bovard2017,Vincent2020,George2020,Kullmann:2021gvo}. 
However, the exact electron fraction ($Y_e$) distribution of this component still vary substantially in different works and require further improved treatment of weak interactions in simulations.}.

The above conclusion reached by \cite{Cote2021} is based on the result that for the typical values of frequency of various \textsl{r}-process sources, the probability of more than one source contributing to $^{129}$I and $^{247}$Cm in the ESS is negligible based on a stochastic  chemical evolution model. 
In this paper we use the turbulent gas diffusion formalism similar to \citet{hotokezaka2015} to simulate the evolution of \textsl{r}-process isotopes at the birth location of the Sun. 
We find that the ``last'' major \textsl{r}-process event only accounts for $\sim 50$--$75\%$ of the total $^{129}$I and $^{247}$Cm measured in the ESS and at least three events are required to account for $\gtrsim 95\%$. 
The minor contributing events can dramatically affect the $^{129}$I/$^{247}$Cm ratio when there are more than one \textsl{r}-process source with distinct production ratios for $^{129}$I/$^{247}$Cm (corresponding to different levels of neutron richness).  
Consequently, we find that the measured meteoritic value may not  
correspond to the value due to a single ``last'' \textsl{r}-process event. Surprisingly, we find that the $^{129}$I/$^{247}$Cm ratio can nevertheless provide important constraints on the neutron richness of \textsl{r}-process when it is combined with the  
meteoritic measurement for the $^{129}$I/$^{127}$I and $^{247}$Cm/$^{235}$U ratio in the ESS.

\section{Method}
In order to calculate the abundance evolution of \textsl{r}-process elements, the frequency of \textsl{r}-process events as well as their locations are
required. The frequency of an \textsl{r}-process source depends on the star formation history (SFH) of the Milky Way. We use the \textsc{omega+} code along with the parameters for the best-fit Milky Way model of \cite{cote2019} to calculate the SFH and the resulting CCSN rate and mass loss factor $f(t)$ (see Appendix~\ref{appA}). 
The rate of \textsl{r}-process events like BNSMs as well as the rate of formation of neutron star binaries (NSBs) are derived by assuming a fraction $f_r$ of CCSNs result in the formation of NSBs. 
We adopt a fiducial value of $f_r=8\times 10^{-4}$ that corresponds to 
a merger frequency $\nu_0\sim 10$~Myr$^{-1}$ at the present time.

We simulate multiple realisations of \textsl{r}-process events in the Milky Way.
For each realisation, the birth times for binaries, which lead to BNSM like events, are generated from a probability distribution 
proportional to the NSB birth rate calculated using \textsc{omega+}. 
The cylindrical radial coordinates for the birth locations in the MW are generated according to a distribution $\propto R\exp(-R/R_d)$, where $R_d$ is the radial scale length for the surface SFR density. In order to account for the inside-out formation of the MW, we adopt time dependent $R_d$ from \citet{Schonrich17}. The corresponding vertical heights are generated according to the distribution of the estimated molecular and atomic gas density in the local solar neighbourhood by \citet{mckee2015}. The merger time $t_{\rm merge}$ for each NSB is sampled from a DTD $\propto t^{-1}$ with a minimum and maximum delay time of 10 Myr and 10 Gyr, respectively. Each NSB is assigned a kick velocity $\vec v_{\rm kick}$ at the time of its birth. The magnitude of $\vec v_{\rm kick}$ is generated from a distribution 
$\propto \exp(-v_{\rm kick}/v_0)$ with a fiducial value of $30\,{\rm km\,s}^{-1}$, whereas the direction of the kick is generated from a uniform isotropic distribution. In order to find the final location of the NSB at the time of its merger, we use \textsc{galpy} \citep{galpy} to trace the motion of the NSB under the influence of the Galactic potential as described in \citet{bwy2020}. The birth time and location for MRSN 
or collapsar like events are generated similar to the NSBs as described above. In this case, however, there are no delays or offsets due to natal kicks.  

In order to calculate the evolution of \textsl{r}-process elements at the Solar location, we use the turbulent gas diffusion formalism by \citet{hotokezaka2015}. The number density of an isotope at a location $\vec r$ and time $t$ is given by 
\begin{equation}
    n(\vec r,t)=\sum_{t_j<t} \frac{Y_{r,j} e^{-f(t)\Delta t_j}}{K_j(\Delta t_j)} \exp\left[-\frac{\left| \vec r-\vec r_j\right|^2}{4D\Delta t_j} -\frac{\Delta t_j}{\tau}\right],
\end{equation}
where $t_j$ is the occurrence time for the $j$th \textsl{r}-process event, $\Delta t_j=t-t_j$, $Y_{r,j}$ is the  number of atoms of the isotope produced by the $j$th  
\textsl{r}-process event, $f(t)$ is the time-dependent loss factor due to star formation and galactic outflows calculated from Milky Way model using \textsc{omega+}, $\tau$ is the lifetime of the isotope. $D$ is the turbulent diffusion coefficient, and $K_j(\Delta t_j)$ is given by 
\begin{equation}
    K_j(\Delta t_j)=\mathrm{min}[(4\pi D \Delta t_j)^{3/2},8\pi h_z D \Delta t_j],
\end{equation}
where $h_z =0.3$~kpc is the vertical scale height. 

The overall mixing of metals depend on the parameters $D$ and the frequency $\nu$ of \textsl{r}-process events with typical mixing timescale $\tau_{\rm mix}$ given by \citep{hotokezaka2015}
\begin{equation}
    \tau_{\rm mix}\approx 200 \left(\frac{\nu}{30~{\rm Myr}^{-1}} \right)^{-2/5}  \left(\frac{D}{0.1~{\rm kpc^2~Gyr^{-1}}} \right)^{-3/5}~{\rm Myr}
    \label{eq:taumix}
\end{equation}
A recent study by \citet{beniamini2020}  
found that in order to satisfy various independent observational constraints such as scatter of stable \textsl{r}-process elements, highest observed \textsl{r}-process enrichment in the solar neighbourhood, as well as constraints from radioactivity in the ESS, it requires that $D\gtrsim 0.1~{\rm kpc^2~Gyr^{-1}}$ and $\nu\lesssim 40~{\rm Myr}^{-1} $ with typical mixing timescale of $ \tau_{\rm mix}\approx 200$ Myr. This also means that $D$ and $\nu$  are related to each other by 
\begin{equation}
    D\approx 0.3 \left (\frac{\nu}{10~{\rm Myr}^{-1}}\right)^{-2/3}~{\rm kpc^2~Gyr^{-1}}.
    \label{eq:D}
\end{equation} 
Because $ \tau_{\rm mix}$ is a fixed parameter, varying the value of $\nu$ is always associated with a corresponding change in $D$. Thus, it is sufficient to consider a single value of $\nu$. 
In our Milky Way model, the current rate of $\nu_0=10~{\rm Myr^{-1}}$ corresponds to a rate of \textsl{r}-process events of  $\nu \approx 15~{\rm Myr^{-1}}$ at the time of SS formation of $t_\odot \sim 9.2$~Gyr. We adopt values of $D=0.1$ and $0.3~{\rm kpc^2~Myr^{-1}}$ that covers values of $D$ 
slightly above and below the corresponding values given by Eq.~\ref{eq:D} for $\nu \approx 15~{\rm Myr^{-1}}$ and correspond to $\tau_{\rm mix}\approx 140$-- $260$ Myr.  

\begin{table*}
\caption{List of important parameters used in this work.} 
\centering 
\begin{tabular}{l c c c}
%\begin{tabular}{|p{2.0cm}|p{2.0cm}|p{2.0cm}|p{2.0cm}|p{2.0cm}|p{2.0cm}|p{2.0cm}|}
%\hline\hline 
\hline
Parameter & Definition & Values &Unit\\
\hline
$D$ & Turbulent diffusion coefficient& 0.1, 0.3 &${\rm kpc^2~Myr^{-1}}$\\
$\nu$ &Total \textsl{r}-process frequency during SS formation&$\sim 15$ & ${\rm Myr}^{-1}$\\
$\nu_{LM}$ &Ratio of frequency of sources $S_L$ to $S_M$ &1&--\\
$\nu_{LH}$ &Ratio of frequency of sources  $S_L$ to $S_H$  &1&--\\
$M_{LM}^{\rm ej}$ &Ratio of ejecta masses of sources $S_L$ to $S_M$  &1/3,1,3&--\\
$M_{LH}^{\rm ej}$ &Ratio of ejecta masses of sources $S_L$ to $S_H$ &1/3,1,3&--\\
$\lambda_{L}$ &$^{129}$I/$^{247}$Cm production ratio for $S_L$  &10&--\\
$\lambda_{L}$ &$^{129}$I/$^{247}$Cm production ratio for $S_M$  &100&--\\
$\lambda_{H}$ &$^{129}$I/$^{247}$Cm production ratio for $S_H$  &1000&--\\
\hline
\end{tabular}
\label{tab:parameters} 
\end{table*}
We compute the number abundance of \textsl{r}-process isotopes using the diffusion prescription discussed above at the solar radius $R_\odot$ at the time when the gas decoupled from the inter stellar medium (ISM) at $t=t_{\rm iso}$ to form the SS. In this work, we consider three different types of \textsl{r}-process sources $S_L$, $S_M$, and $S_H$ defined by distinct $^{129}$I/$^{247}$Cm production ratios of $\lambda_L$, $\lambda_M$, and $\lambda_H$. The subscripts $L$, $M$, and $H$ refer to low, medium and high values, respectively, for the $^{129}$I/$^{247}$Cm production ratio corresponding to different astrophysical sites with varying neutron richness. We adopt fiducial values of  $\lambda_L=10$, $\lambda_M=100$, and $\lambda_H=1000$. The adopted values roughly correspond to values expected from neutron rich dynamical ejecta during a BNSM or NS--BH merger ($\lambda_L$), moderately neutron rich disk ejecta following BNSMs ($\lambda_M$), and a low neutron rich ejecta from MRSN events ($\lambda_H$) from theoretical models reported in \citet{Cote2021}. For each \textsl{r}-process event the isotopic production ratio is taken to be $^{129}$I/$^{127}$I= 1.46, $^{247}$Cm/$^{235}$U= 0.20, $^{244}$Pu/$^{238}$U= 0.40, and $^{235}$U/$^{238}$U= 1.05. 
These values for actinide ratios are generally consistent with the predictions from \citet{Mendoza2015} and 
 \citet{Wu:2016pnw} which computed the \textit{r}-process nucleosynthesis in the BNSM dynamical ejecta and in the BH--accretion disk outflows using different nuclear physics inputs. The value for $^{129}$I/$^{127}$I ratio corresponds to the solar \textsl{r}-process value adapted from \citet{Sneden2008}. We also account for the contribution of \textsl{s}-process to the ESS value of $^{127}$I by multiplying the a factor 1.06 which is consistent the solar abundance decomposition by \citet{Sneden2008}.

We consider three scenarios, where only two out of the three different types of \textsl{r}-process sites contribute. Additionally, we consider the scenario where all three types of sources contribute. 
The relevant parameters that impact the evolution of isotopic ratios of SLRs produced by \textsl{r}-process are the frequency $\nu_i$ of each \textsl{r}-process source, the corresponding production ratio $\lambda_i$, the relative ratio of the ejected mass $M^{\rm ej}_{ij}=M^{\rm ej}_i/M^{\rm ej}_j$ for sources $S_i$ and $S_j$, and the value of the diffusion coefficient $D$. We list the definitions and the adopted values of relevant model parameters in Table~\ref{tab:parameters}.

\begin{figure*}
\centerline{\includegraphics[width=\textwidth]{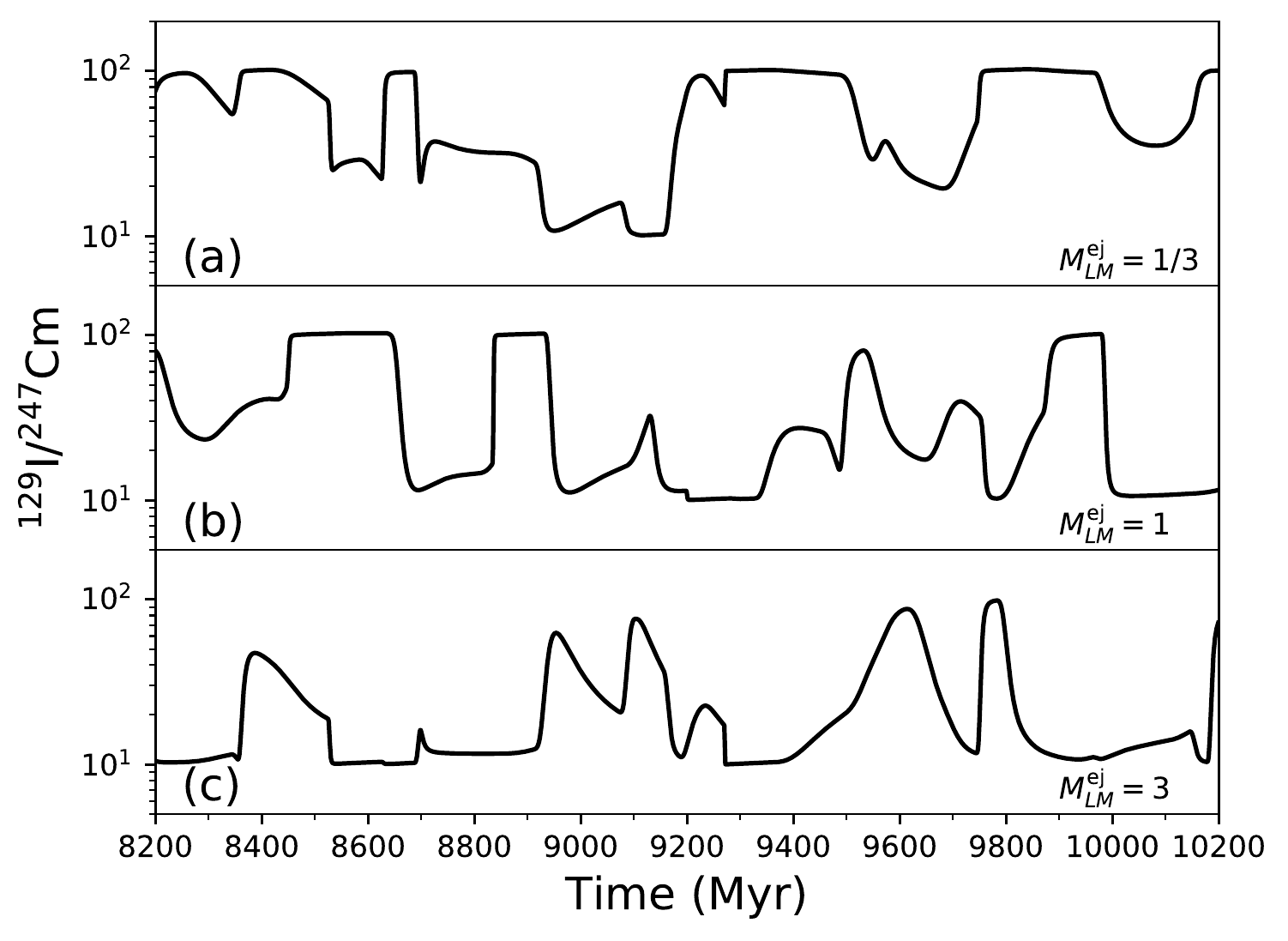}}
\caption{Evolution of $^{129}$I/$^{127}$Cm at the solar location for models with two equally frequent ($\nu_{LM}=1$) sources $S_L$ and $S_M$ with $\lambda_L=10$ and $\lambda_M=100$ with $D=0.1$~kpc$^2$ Gyr$^{-1}$ for three different values of $M^{\rm ej}_{LM}$.}

\label{fig:I129_Cm247_Dp1}
\end{figure*}
 
\section{Results}
We first consider two \textsl{r}-process sources $S_L$ and $S_M$ that are BNSM like.
We assume that both of them are equally frequent, i.e., $\nu_{LM}=\nu_L/\nu_M=1$ with $\nu_{L}+\nu_{M}=\nu$, and have the same kick velocity distribution.
For their relative ejecta mass ratios, we consider three different values of
$M^{\rm ej}_{LM}=1/3,1,$ and $3$.
$M^{\rm ej}_{LM}=1$ corresponds to the scenario where both $S_L$ and $S_M$ contribute equally to the main \textsl{r}-process. The other two values of $M^{\rm ej}_{LM}$ correspond to scenarios where one of the sources is the dominant site for the main \textsl{r}-process.
Figure~\ref{fig:I129_Cm247_Dp1} shows the evolution of $^{129}$I/$^{247}$Cm ratio at the solar location for $D=0.1$~kpc$^2$ Gyr$^{-1}$ with three different values of $M^{\rm ej}_{LM}$ for one single representative realisation. As expected, the value of the ratio varies between $\lambda_L$ and $\lambda_M$. However, the ratio not only takes extremal values, but also assumes values of $\sim 30-70$ for substantial lengths of time. This occurs even when the ejecta mass is higher for one of the sources i.e, $M^{\rm ej}_{LM}=1/3$ and $3$.

% \begin{table*}
% \caption{Probability distribution of $^{129}$I/$^{247}$Cm ratio in the ESS for criteria T0 (see text) for two equally frequent ($\nu_{LM}=1$) \textsl{r}-process sources with $\lambda_{L}=10$, $\lambda_{M}=100$. All models have $\nu_0=10$~Myr$^{-1}$
% %$r_0=10$ Myr$^{-1}$,  $\nu_L=\nu_M=0.5$, $\lambda_{L}=10$, $\lambda_{\mrw{M}}=100$, 
% and $v_{\rm kick}^0=30$ km s$^{-1}$. Unit for $D$ is kpc$^2$ Gyr$^{-1}$.} 
% \centering 
% %\begin{tabular}{c c c c c c c} 
% \begin{tabular}{|p{3cm}|p{1cm}|p{1cm}|p{1cm}|p{1cm}|p{1cm}|p{1cm}|p{1cm}|}
% \hline\hline 
% Model Parameters&\multicolumn{7}{c|}{Probability of ESS $^{129}$I/$^{247}$Cm Ratio within an interval}\\ [0.5ex]
% \hline
% $(D,M^{\rm ej}_{LM})$ & 10-20 & 20-30 & 30-50 & 50-70 & 70-80 & 80-90 & 90-110 \\ [0.5ex] 
% \hline 
% (0.3,1/3) &0.41 &0.10& 0.11&0.09 &0.05&0.07& 0.17\\
% (0.1,1/3) &0.45 &0.06& 0.07&0.06 &0.03&0.04& 0.30\\
% \hline
% (0.3,1)&  0.59 &0.11&0.09&0.07 &0.03&0.04& 0.09\\
% (0.1,1)&  0.65 &0.04& 0.05&0.04 &0.02&0.03& 0.17\\

% \hline
% (0.3,3)&  0.74 &0.08&0.07&0.05 &0.02&0.02& 0.03\\
% (0.1,3)&  0.77 &0.04&0.03 &0.04 &0.02&0.02& 0.09\\

% \hline 
% \end{tabular}
% \label{tab:minimum} 
% \end{table*}

\begin{table*}
\caption{Probability distribution of $^{129}$I/$^{247}$Cm ratio in the ESS for criteria T0 (see text) for two equally frequent \textsl{r}-process sources $S_L$ and $S_M$ with $\lambda_{L}=10$, $\lambda_{M}=100$. 
Unit for $D$ is kpc$^2$ Gyr$^{-1}$.} 
\centering 
\begin{tabular}{c c c c c c c c c c} 
%\begin{tabular}{|p{3cm}|p{1cm}|p{1cm}|p{1cm}|p{1cm}|p{1cm}|p{1cm}|p{1cm}|}
\hline\hline 
Model Parameters&\multicolumn{7}{c|}{Probability of ESS $^{129}$I/$^{247}$Cm Ratio within an interval}\\ [0.5ex]
\hline
$(D,M^{\rm ej}_{LM})$ & 10--20 & 20--30 & 30--40&40--50 &50--60& 50--70 & 70--80 & 80--90 & 90--110 \\ [0.5ex] 
\hline 
(0.3,1/3) &0.41 &0.11 &0.07 &0.04 &0.04 &0.04 &0.05&0.08& 0.18\\
(0.1,1/3) &0.45 &0.06 &0.04 &0.03 &0.03 &0.03 &0.03&0.04& 0.30\\
\hline
(0.3,1)&  0.56 &0.09 &0.06 &0.05 &0.04 &0.03  &0.03&0.04& 0.09\\
(0.1,1)&  0.63 &0.06 &0.04 &0.03 &0.01 &0.02 &0.02&0.03& 0.17\\

\hline
(0.3,3)&  0.75 &0.07&0.04 &0.03 &0.02 &0.02 &0.02&0.02& 0.03\\
(0.1,3)&  0.75 &0.04&0.03 &0.02 &0.02 &0.01 &0.02&0.03& 0.10\\

\hline 
\end{tabular}
\label{tab:minimum} 
\end{table*}

\subsection{Results with Minimum Criteria}

The formation of meteorites in the ESS is always preceded with some isolation time. This is because there is always a interval $\Delta_{\rm iso}$ between the time when the molecular cloud, from which the SS formed, decouples from the ISM at $t_{\rm iso}$ and the formation of the SS at $t_\odot=t_{\rm iso}+\Delta_{\rm iso}$. During this time, the gas from which the SS was formed does not receive any contribution for \textsl{r}-process isotopes and the SLRs free decay for a period of $\Delta_{\rm iso}$.
Thus, in order for the isotope ratio $^{129}$I/$^{127}$I and $^{247}$Cm/$^{235}$U to be a viable candidate for the measured value at ESS formation time, 
the SLR isotopic ratio resulting from  chemical evolution model has to be greater than the ESS meteoritic values. Because SLR $^{244}$Pu is also produced exclusively by \textsl{r}-process, this also applies to the  $^{244}$Pu/$^{238}$U ratio.  
We define this as the minimum criteria and  
refer to this criteria as T0 with the adopted
observed ESS abundance of $^{129}$I/$^{127}$I$= 1.28\times 10^{-4}$ \citep{Ott2016}, $^{247}$Cm/$^{235}$U$= 5.6\times 10^{-5}$ \citep{Tang+2017}, and $^{244}$Pu/$^{238}$U$= 7\times 10^{-3}$ \citep{Hudson1989}. 
We simulate 1000 realisations of \textsl{r}-process enrichment in the Milky Way that satisfy criteria T0. This gives us a distribution of ESS value of the $^{129}$I/$^{247}$Cm ratio  that ranges roughly from $\lambda_L$ to $\sim \lambda_M$. 
The distribution allows us to directly calculate the probability of finding the ESS $^{129}$I/$^{247}$Cm ratio within a given interval between $\lambda_L$ and $\sim \lambda_M$. 
Table~\ref{tab:minimum} shows the results using $\lambda_M=100$ and $\lambda_L=10$ for three different choices of $M^{\rm ej}_{LM}$, and two different values of $D$. Interestingly, for all cases, the probability always peaks at values close to $\lambda_L$.
In all cases with $D=0.3$ kpc$^2$ Gyr$^{-1}$, there is a $\gtrsim 12$--$20\%$ probability of getting intermediate $^{129}$I/$^{247}$Cm values of $30$--$70$ that is comparable to the probability of getting values close to $\lambda_M$. When $D=0.1$ kpc$^2$ Gyr$^{-1}$, the  probability for intermediate values are somewhat lower but is still comparable to the probability for getting values close to $\lambda_M$.

In order to understand the probability distribution of the  $^{129}$I/$^{247}$Cm ratio,  
the history of enrichment of \textsl{r}-process SLRs at the solar location needs to be analysed. In general, the abundance of \textsl{r}-process isotopes at the solar location $R_\odot$ at $t_\odot$ receives contribution from all  \textsl{r}-process events that has occurred before $t_{\rm iso}$. However, for SLRs,  only a handful of events that has occurred at $t'$ where $t_{\rm iso}-t'$ is within a few lifetimes of the SLRs at locations relatively close to the Sun can contribute. For any isotope including SLRs, the contribution from all the past events at a particular $(R_\odot,t)$ can be sorted in terms of the fraction contributed to the total amount. It is useful to define the quantity $f^{\rm iso}_{\rm h1}$ as the highest fraction of the ESS value contributed by a single event for a particular isotope. This is the same as $f_{\rm last}$ defined by \citet{beniamini2020}. We can further define $f^{\rm iso}_{\rm h2}$ and  $f^{\rm iso}_{\rm h3}$ as the fractions contributed by the second  and third highest contributors for a particular isotope, respectively. Table~\ref{tab:h1h2h3_mimimum} shows the values of $f_{\rm h1}$, $f_{\rm h2}$, and $f_{\rm h3}$ for $^{129}$I corresponding to the models listed in Table~\ref{tab:minimum}. As can be seen from the table, on average, the single major contributor accounts for only $70$--$84\%$  and at least three events are required to account for $\gtrsim 95\%$ of the total $^{129}$I in the ESS.

\begin{table*}
\caption{The mean and median values of  $f_{\rm h1}$, $f_{\rm h2}$, and $f_{\rm h3}$ along with the 95th percentile range for $^{129}$I for the models listed in Table~\ref{tab:minimum}. Unit for $D$ is kpc$^2$ Gyr$^{-1}$.} 
\centering 
\begin{tabular}{c c c c c c} 
%\begin{tabular}{|p{2.0cm}|p{2.0cm}|p{2.0cm}|p{2.0cm}|p{2.0cm}|p{2.0cm}|}
%\hline\hline 
%Model Parameters&\multicolumn{3}{c|}{ }\\ [0.5ex]
\hline
$(D,M^{\rm ej}_{LM})$ & &$f^{\rm ^{129}I}_{\rm h1}$ & $f^{\rm ^{129}I}_{\rm h2}$ & $f^{\rm ^{129}I}_{\rm h3}$ & $f^{\rm ^{129}I}_{\rm h1+h2+h3}$\T \B\\ 
\hline
\multirow{3}{*}{(0.3,1/3)} & mean&0.702 &0.167 &0.062 &0.931\\
 &median&0.714   &0.158   &0.039 &0.961\\
 &95th percentile &0.350--0.988 & 0.006--0.394 & 0.002--0.191 & 0.753--0.999 \\
 \hline
\multirow{3}{*}{(0.1,1/3)}  & mean&0.838  &0.119   &0.028 &0.985\\
&median&0.908  &0.062   &0.006 &0.997\\
&95th percentile&0.488--0.999 & 0.000--0.400&0.000--0.128&0.923--1.000\\
\hline
\hline
\multirow{3}{*}{(0.3,1)} & mean&0.701 &0.164   &0.062 &0.927\\
&median&0.705 &0.155   &0.044 &0.959\\
&95th percentile&0.349--0.9991 & 0.005--0.374 & 0.001--0.189 & 0.746--0.999\\
\hline
\multirow{3}{*}{(0.1,1)} & mean&0.823   & 0.127  &0.032 &0.982\\
&median&0.893   & 0.076  &0.008 &0.996\\
&95th percentile&0.497--0.999 & 0.000--0.398 & 0.000--0.142&0.913--1.000\\
\hline
\hline
\multirow{3}{*}{(0.3,3)} & mean&0.719  &0.158  &0.058 &0.936\\
&median&0.753 &0.140  &0.033 &0.970\\
&95th percentile&0.348--0.990 & 0.005--0.384 & 0.001--0.190 &0.772--0.999\\
\hline
\multirow{3}{*}{(0.1,3)} & mean&0.826  &0.124  &0.031 &0.982\\
&median&0.906 &0.068  &0.007 &0.997\\
&95th percentile&0.459--0.999 & 0.000--0.395 & 0.000--0.148 &0.905--1.000\\
\hline
\end{tabular}
\label{tab:h1h2h3_mimimum} 
\end{table*}

The contributions from the second and third highest contributors have a critical impact on the $^{129}$I/$^{247}$Cm ratio. To illustrate this, let us consider the case with $D=0.1$ kpc$^2$ Gyr$^{-1}$ and $M^{\rm ej}_{LM}=1$ where $\approx 99\%$ of the total $^{129}$I is produced by three events with mean values of $f_{\rm h1}=0.84$, $f_{\rm h2}=0.12$, and $f_{\rm h3}=0.03$. In this case both $S_L$ and $S_M$ produce the same amount of $^{129}$I whereas the former produces a factor 10 higher amount of $^{247}$Cm. If $N_0$ denotes the number of atoms of $^{129}$I produced by each source, then the corresponding yield of $^{247}$Cm is $0.1N_0$ and $0.01N_0$ for $S_L$ and $S_M$, respectively. 
For simplicity, let us assume that the three highest contributing events account for all the $^{129}$I. In this case, for these three events $({\rm h1,h2,h3})$, the 8 different possible combinations are $(S_L,S_L,S_L)$, $(S_L,S_L,S_M)$, $(S_L,S_M,S_L)$, $(S_L,S_M,S_M)$, $(S_M,S_M,S_M)$, $(S_M,S_L,S_M)$, $(S_M,S_L,S_L)$, and $(S_M,S_M,S_L)$, which all have equal occurrence probability of $12.5\%$ and produce equal amounts of $^{129}$I ($0.99 N_0$ atoms). 
Summing over the produced $^{247}$Cm atoms gives rise to values of $^{129}$I$/^{247}$Cm equal to 10.1, 10.4, 11.3, 11.7, 101.0, 48.3, 42.7, 79.4 for these 8 scenarios, respectively.
Clearly, the first four combinations ($50\%$ probability) where the dominant $^{129}$I contributor ${\rm h1}=S_L$, the total number of $^{247}$Cm atoms is dominated by the $S_L$ source, leading to $^{129}$I/$^{247}$Cm ratios close to $\lambda_L=10$.
In contrast, in the latter 4 combinations where ${\rm h1}=S_M$, 
except for the case $(S_M,S_M,S_M)$, the total $^{247}$Cm gets substantial contribution from $S_L$ sources from ${\rm h2}$ and ${\rm h3}$ events as at least one them is of $S_L$. 
Thus, in the latter 4 cases, $^{129}$I/$^{247}$Cm ratio is close to $\lambda_M$ only for $(S_M,S_M,S_M)$ with a probability of $12.5\%$.
For the other three combinations, the $^{129}$I/$^{247}$Cm ratio takes intermediate values.

\begin{table*}
\caption{Probability distribution of $^{129}$I/$^{247}$Cm ratio in the ESS for concordance criteria T10 and T20 (see text) for the models listed in Table~\ref{tab:minimum}.
Values with the minimal criteria T0 are the same as in Table~\ref{tab:minimum}. Unit for $D$ is kpc$^2$ Gyr$^{-1}$.} 
\centering 
\begin{tabular}{c c c c c c c c c c c} 
%\begin{tabular}{|p{1.4cm}|p{1.4cm}|p{.9cm}|p{.9cm}|p{.9cm}|p{1cm}|p{.9cm}|p{.9cm}|p{1cm}|}
\hline\hline 
Model& &\multicolumn{7}{c|}{Probability of $^{129}$I/$^{247}$Cm within an interval }\\ [0.5ex]
\hline
$(D,M^{\rm ej}_{LM})$ &Criteria & 10--20 & 20--30 & 30--40& 40--50 & 50--60& 60--70 & 70--80 & 80--90 & 90--110 \\ [0.5ex] 
\hline -
\multirow{3}{*}{(0.3,1/3)} &T0 &0.41 &0.11 &0.07 &0.04 &0.04 &0.04 &0.05&0.08& 0.18\\
&T10 & 0.00 & 0.10 &0.56 &0.32 &0.03 &0.00 & 0.00&0.00&0.00\\
&T20 & 0.00 & 0.24 &0.35 &0.28 &0.13 &0.00 & 0.00&0.00&0.00\\
%&T40 &  0.11 &0.19& \textbf{0.27} & \textbf{0.21}&0.13&0.07& 0.03\\
\hline
\multirow{3}{*}{(0.1,1/3)} & T0 &0.45 &0.06 &0.04 &0.03 &0.03 &0.03 &0.03&0.04& 0.30\\
&T10&  0.00 & 0.10&0.59 &0.29 &0.03 &0.00 & 0.00& 0.00& 0.00\\ 
&T20&  0.00 &0.22 &0.34 &0.27 &0.15 &0.02 & 0.00&0.00 & 0.00\\
%&T40&  0.10 &0.16& 0.22 & 0.19&0.09&0.08& 0.15\\
\hline
\hline
\multirow{3}{*}{(0.3,1)}& T0 &0.56 &0.09 &0.06 &0.05 &0.04 &0.03  &0.03&0.04& 0.09\\
&T10&  0.32 &0.68&0.00 &0.00 &0.00 &0.00 &0.00&0.00& 0.00\\
&T20&  0.46 &0.47&0.07 &0.00 &0.00 &0.00 &0.00&0.00& 0.00\\
%&T40&  0.64 &0.17& \textbf{0.18} & \textbf{0.01}&0.00&0.00& 0.00\\
\hline
\multirow{3}{*}{(0.1,1)}&T0 &0.63 &0.06 &0.04 &0.03 &0.01 &0.02 &0.02&0.03& 0.17\\ 
&T10 & 0.35 & 0.63&0.02 &0.00 &0.00 &0.00 & 0.00& 0.00& 0.00\\
&T20 & 0.50 &0.39 &0.11 &0.00 &0.00 &0.00 & 0.00&0.00& 0.00\\
%&T40&  0.75 &0.13& \textbf{0.11}& \textbf{0.02} &0.00&0.00& 0.00\\
\hline
\hline
\multirow{3}{*}{(0.3,3)}& T0& 0.75 &0.07&0.04 &0.03 &0.02 &0.02 &0.02&0.02& 0.03\\
&T10&  0.98 &0.02&0.00 &0.00 &0.00 &0.00 &0.00&0.00& 0.00\\
&T20&  0.93 &0.07&0.00 &0.00 &0.00 &0.00 &0.00&0.00& 0.00\\
%&T40&  0.83 &0.12& \textbf{0.05} & \textbf{0.00}&0.00&0.00& 0.00\\
\hline
\multirow{3}{*}{(0.1,3)}&T0& 0.75 &0.04&0.03 &0.02 &0.02 &0.01 &0.02&0.03& 0.10\\
&T10&  0.98 &0.02&0.00 &0.00 &0.00 &0.00 &0.00&0.00& 0.00\\
&T20&  0.95 &0.05&0.00 &0.00 &0.00 &0.00 &0.00&0.00& 0.00\\
%&T40&  0.88 &0.09& \textbf{0.03} & \textbf{0.00}&0.00&0.00& 0.00\\
\hline
\end{tabular}
\label{tab:2sourcesLM} 
\end{table*}

\begin{figure*}
\centerline{\includegraphics[width=\textwidth]{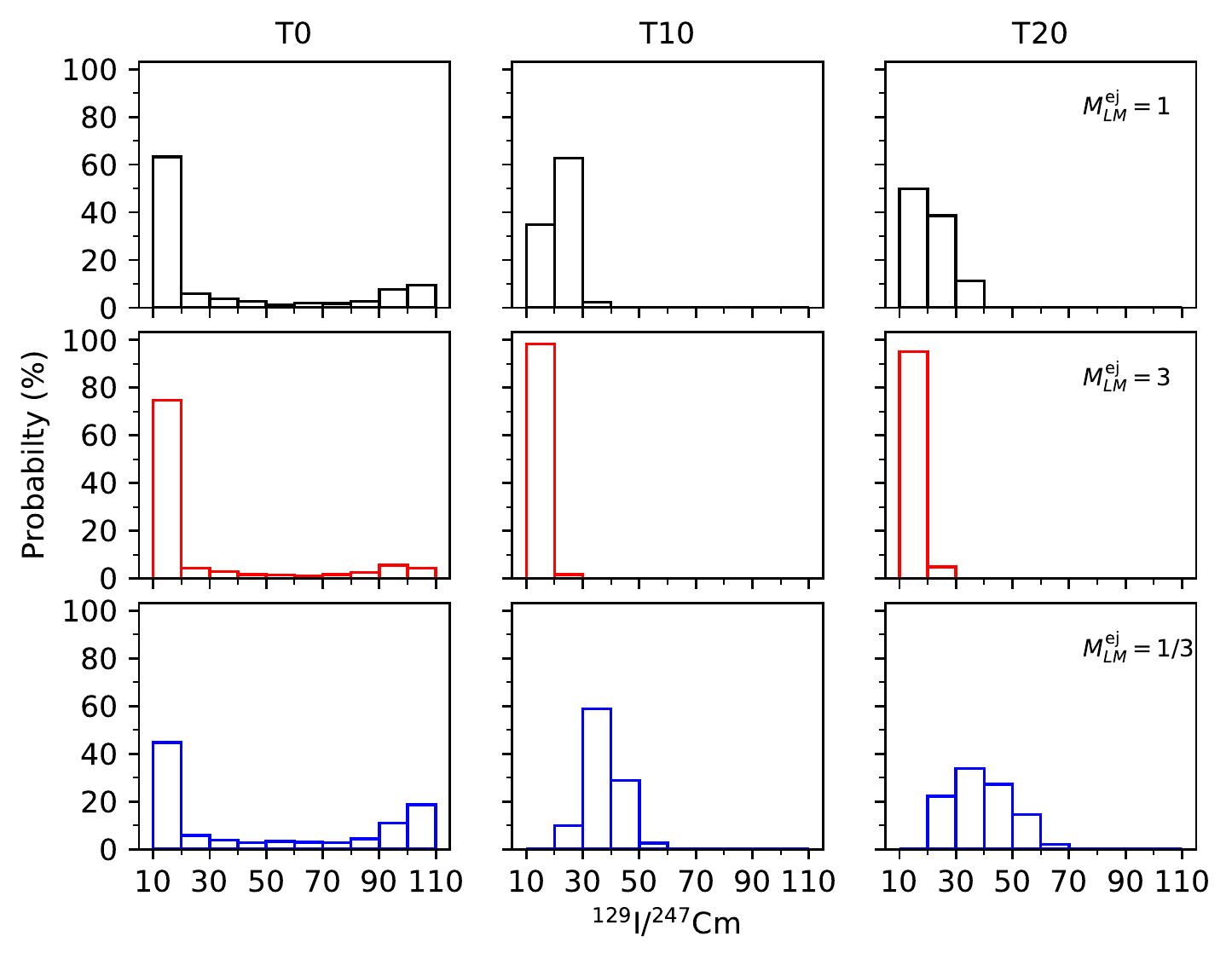}}
\caption{Probability distribution of $^{129}$I/$^{127}$Cm ratio in the ESS for two equally frequent \textsl{r}-process sources $S_L$ and $S_M$ with $\lambda_L=10$ and $\lambda_M=100$ for  $M_{LM}^{\rm ej}=1$ (black), 3 (red), and 1/3 (blue) and tolerances T0 (left vertical panel), T10 (middle vertical panel), and T20 (right vertical panel) corresponding to the values listed in Table~\ref{tab:2sourcesLM}. All models have $D=0.1$ kpc$^2$ Gyr$^{-1}$.}

\label{fig:Dp1_LM_hist}
\end{figure*}

The probabilities estimated from above simple illustration using the mean values of $f_{\rm h1}$, $f_{\rm h2}$, and $f_{\rm h3}$ agrees qualitatively with the values listed Table~\ref{tab:minimum}. The quantitative differences is due to the fact that the mean value does not fully represent the distribution of $f_{\rm h1}$, $f_{\rm h2}$, and $f_{\rm h3}$. This is particularly true for models with lower value of $D=0.1$ kpc$^2$ Gyr$^{-1}$ which has distribution with a long tail, as evident from the large 95th percentile ranges along with large differences between the  mean and median values.
Nevertheless, the simple analysis clearly illustrates the fact that $^{247}$Cm can receive substantial contribution from the subdominant ${\rm h2}$ and ${\rm h3}$ events and directly impacts the $^{129}$I/$^{247}$Cm ratio. In particular, this shows why the probability distribution peaks at values closer to $\lambda_L$ rather than $\lambda_M$ and why it also takes  
intermediate values. 

\begin{table}
\caption{The mean values of  $f_{\rm h1}$, $f_{\rm h2}$, and $f_{\rm h3}$ for $^{129}$I for the models listed in Table~\ref{tab:minimum}.
Values with the minimal criteria T0 are the same as in Table~\ref{tab:h1h2h3_mimimum}. Unit for $D$ is kpc$^2$ Gyr$^{-1}$.}
\centering 
\begin{tabular}{c c c c c c} 
%\begin{tabular}{|p{2.0cm}|p{2.0cm}|p{2.0cm}|p{2.0cm}|p{2.0cm}|p{2.0cm}|}
%\hline\hline 
%Model Parameters&\multicolumn{3}{c|}{ }\\ [0.5ex]
\hline
$(D,M_{LM}^{ej})$& Criteria &$f^{\rm ^{129}I}_{\rm h1}$ & $f^{\rm ^{129}I}_{\rm h2}$ & $f^{\rm ^{129}I}_{\rm h3}$ & $f^{\rm ^{129}I}_{\rm h1+h2+h3}$\T \B\\ 
\hline
\multirow{3}{*}{ (0.3,1/3)} & T0 &0.702 &0.167 &0.062 &0.931\\
 &T10 &0.693 &0.196 &0.071 &0.960\\
 &T20 &0.591 &0.196 &0.094 &0.882\\
 \hline
\multirow{3}{*}{ (0.1,1/3)} & T0 & 0.838&0.119 & 0.028&0.985\\
 &T10 &0.690 &0.198 & 0.073&0.961\\
 &T20 &0.691 &0.198 &0.070&0.960\\
 \hline
 \hline
 \multirow{3}{*}{ (0.3,1)} & T0 &0.701 &0.164 &0.062 &0.927\\
 &T10 &0.458 &0.292 &0.116 &0.865\\
 &T20 &0.483 &0.273 &0.111 &0.868\\
 \hline
\multirow{3}{*}{ (0.1,1)} & T0 &0.823 &0.127 &0.032 &0.982\\
 &T10 &0.525 & 0.340& 0.087&0.952\\
 &T20 &0.561 &0.311& 0.085&0.957\\
 \hline
 \hline
  \multirow{3}{*}{ (0.3,3)} & T0 &0.719 &0.158 &0.058 &0.936\\
 &T10 & 0.639&0.249 &0.072 &0.959\\
 &T20 & 0.626&0.200 &0.079 &0.905\\
 \hline
\multirow{3}{*}{ (0.1,3)} & T0 &0.826 &0.124 &0.031 &0.982\\
 &T10 &0.656 &0.246 & 0.063&0.965\\
 &T20 &0.768 &0.168 &0.040 &0.977\\
\hline

\hline
\end{tabular}
\label{tab:h1h2h3_concordance} 
\end{table}
\subsection{Results with Concordant Decay Interval}\label{sec:concordant}
So far, we have considered Milky Way realisations that satisfy the minimum criteria T0, namely, that the ratios of $^{129}$I/$^{127}$I, $^{247}$Cm/$^{235}$U, and $^{244}$Pu/$^{238}$U are higher than the mean measured values in the ESS. This, however, does not guarantee that the ratios are actually compatible with the measurements. As mentioned before, after the star forming gas in the molecular or giant molecular cloud decouples from the ISM, each SLR undergoes free decay for the same length of isolation time $\Delta_{\rm iso}$ before the formation of meteorites in the ESS \citep{wasserburg2006,lugaro2018}. The value of $\Delta_{\rm iso}$ can be directly calculated for each SLR separately from the relation
\begin{equation}
    \left( \frac{N_R}{N_I}\right )_{\rm ESS}\simeq\left( \frac{N_R}{N_I}\right )_{\rm iso}e^{-\Delta_{\rm iso}/\tau_R},
    \label{eq:Delta}
\end{equation}
where $(N_R/N_I)_{\rm ESS}$ is the meteoritic ratio of the SLR $R$ with respect to the stable or long-lived isotope $I$ and $(N_R/N_I)_{\rm iso}$ is the corresponding ratio when the star forming gas decouples from the ISM at $t_{\rm iso}$\footnote{The result in Eq.~\ref{eq:Delta} holds exactly when the long-lived isotope $I$ is stable or when the SLR $R$ does not decay to $I$.}. In order for the ratios to be compatible with the ESS measurements,  the value of $\Delta_{\rm iso}$ calculated for $^{129}$I/$^{127}$I, $^{247}$Cm/$^{235}$U, and  $^{244}$Pu/$^{238}$U need to match. This criteria, however, is not meaningful as the probability for the three different values of $\Delta_{\rm iso}$ to be exactly identical is zero. We can however, define compatibility within some tolerance range about the mean measured value of $(N_R/N_I)_{\rm ESS}$. In this case, for each of the three SLR to stable isotope ratios, we get a range of $\Delta_{\rm iso}=(\Delta^{\rm min}_{\rm iso},\Delta^{\rm max}_{\rm iso})$.
We consider a realisation to be compatible (concordant) with the ESS measurements if the range of the three $\Delta_{\rm iso}$ values overlap.

\begin{table*}
\caption{Probability distribution of $^{129}$I/$^{247}$Cm ratio in the ESS for concordance criteria T0, T10 and T20 (see text) for two equally frequent \textsl{r}-process sources: one MRSN like $S_H$ source with $\lambda_H=1000$ and one BNSM like $S_L$ source with $\lambda_L=10$.} 
\centering 
\begin{tabular}{c c c c c c c c c c c c} 
%\begin{tabular}{|p{2.0cm}|p{1.4cm}|p{.9cm}|p{.9cm}|p{.9cm}|p{1cm}|p{.9cm}|p{.9cm}|p{1cm}|p{1cm}|p{1cm}|p{.9cm}}
\hline\hline 
Model& &\multicolumn{10}{c|}{Probability of $^{129}$I/$^{247}$Cm within an interval }\\ [0.5ex]
\hline
$(D,M^{\rm ej}_{LH})$ &Criteria & 10-20 & 20-30 & 30-50 &50-70 & 70-80 & 80-90 & 90-110 &110-130&130-150& $>150$\\ [0.5ex] 
\hline 
\multirow{3}{*}{(0.1,1)}  &T0 & 0.63& 0.06& 0.05& 0.03& 0.01& 0.01& 0.02& 0.01& 0.01& 0.17\\
&T10 & 0.04& 0.69& 0.27& 0.00& 0.00&  0.00& 0.00& 0.00& 0.00& 0.00   \\
&T20 & 0.16& 0.49& 0.34& 0.01& 0.00&  0.00& 0.00& 0.00& 0.00& 0.00   \\
%&T40 & 0.55&0.19& 0.18& 0.07& 0.01& 0.00& 0.00& 0.00& 0.00& 0.00\\
\hline
\multirow{3}{*}{(0.3,1)}  &T0 &0.53& 0.09& 0.08& 0.04& 0.02& 0.01& 0.02& 0.01& 0.02&  0.17\\
&T10 & 0.03& 0.84& 0.13& 0.00& 0.00&  0.00& 0.00& 0.00& 0.00& 0.00  \\
&T20 & 0.21& 0.50& 0.29& 0.01& 0.00&  0.00& 0.00& 0.00& 0.00& 0.00  \\
%&T40 & 0.54& 0.20&  0.18& 0.07& 0.00& 0.00& 0.00& 0.00& 0.00& 0.00\\
\hline
\hline
\multirow{3}{*}{(0.1,1/3)}   &T0 & 0.47& 0.05& 0.06&  0.03& 0.02& 0.01& 0.02& 0.01& 0.01& 0.32\\
&T10 &0.00 &0.00& 0.25& 0.60& 0.10& 0.04& 0.01& 0.00& 0.00& 0.00\\
&T20 &0.00 &0.00& 0.33& 0.42& 0.12& 0.07& 0.05& 0.00& 0.00& 0.00\\
%&T40 & 0.01& 0.13& 0.30& 0.21& 0.07& 0.07& 0.12& 0.07& 0.03& 0.01\\
\hline
\multirow{3}{*}{(0.3,1/3)}   &T0 & 0.37& 0.08& 0.09& 0.06& 0.02& 0.01& 0.03& 0.02& 0.01& 0.31 \\
&T10 & 0.00& 0.10& 0.88& 0.03& 0.00& 0.00& 0.00& 0.00& 0.00&  0.00 \\
&T20 & 0.00& 0.00& 0.45& 0.41& 0.10& 0.03& 0.00& 0.00& 0.00&  0.00 \\
%&T40 & 0.00& 0.14&  0.31& 0.21& 0.09& 0.08& 0.10& 0.06& 0.01& 0.00\\
\hline
\hline
\multirow{3}{*}{(0.1,3)}   &T0 & 0.76& 0.04& 0.05& 0.02&  0.01& 0.00& 0.01& 0.01& 0.00& 0.10\\
&T10 &0.98& 0.02& 0.00& 0.00& 0.00& 0.00& 0.00&  0.00& 0.00& 0.00\\
&T20 &0.88& 0.12& 0.00& 0.00& 0.00& 0.00& 0.00&  0.00& 0.00& 0.00\\
%&T40 &0.87& 0.09& 0.04& 0.00& 0.00& 0.00& 0.00&  0.00& 0.00& 0.00\\
\hline
\multirow{3}{*}{(0.3,3)}   &T0 &0.70& 0.07& 0.06& 0.03& 0.01& 0.01& 0.01& 0.01& 0.01& 0.09\\
&T10 &0.96& 0.04& 0.00& 0.00& 0.00& 0.00& 0.00&  0.00& 0.00& 0.00\\
&T20 &0.87& 0.13& 0.00& 0.00& 0.00& 0.00& 0.00&  0.00& 0.00& 0.00\\
%&T40 &0.77& 0.17& 0.07& 0.00& 0.00& 0.00& 0.00&  0.00& 0.00& 0.00\\
\hline
\end{tabular}
\label{tab:2sourcesLH} 
\end{table*}
We adopt tolerance levels of $10\%$ (criteria T10) and $20\%$ (criteria T20) about the mean measured meteoritic value in the ESS for each of the three ratios. Table~\ref{tab:2sourcesLM} shows the %results for the 
resulting probabilities for the ESS $^{129}$I/$^{247}$Cm ratio in different intervals between $\sim \lambda_L$ to $\lambda_M$. 
When concordance is imposed with the adopted tolerances, the probability distributions are dramatically different compared to case where only the minimum criteria (T0) is imposed (see Fig.~\ref{fig:Dp1_LM_hist}). For both concordance criteria, the probability distribution for the $^{129}$I/$^{247}$Cm ratio is $\lesssim 6\lambda_L$ for all models. In fact, for $M^{\rm ej}_{LM}=1$ and $3$, the $^{129}$I/$^{247}$Cm ratio is limited to values $\leq 4\lambda_L$. 
Only the models with $M^{\rm ej}_{LM}=1/3$ have non-negligible probability of up to 17\% for $^{129}$I/$^{247}$Cm ratio to be larger than $>5 \lambda_L$. 
Thus, remarkably, even in the case where $\lambda_M$ is the major contributor to \textsl{r}-process, i.e., $M^{\rm ej}_{LM}=1/3$, the probability for the $^{129}$I/$^{247}$Cm ratio to be close to $\lambda_M$ is zero. As we can see from Table~\ref{tab:2sourcesLM}, in all models with T10 and T20 criteria, the probability is strongly peaked at values either close to $\lambda_L$ or at intermediate values of $\sim 3-5\lambda_L$. Models with $M^{\rm ej}_{LM}=3$ i.e., $S_L$ as the dominant \textsl{r}-process source, where the ratio is almost entirely limited to values close to $\lambda_L$, are the only ones where the measured $^{129}$I/$^{247}$Cm ratio corresponds to the true production ratio of one of the sources with almost 100\% probability. In all other cases, the ESS value of $^{129}$I/$^{247}$Cm can only provide constraints on the range spanned by $\lambda_L$ and $\lambda_M$ but cannot directly constrain their exact values.    
\begin{figure*}
\centerline{\includegraphics[width=\textwidth]{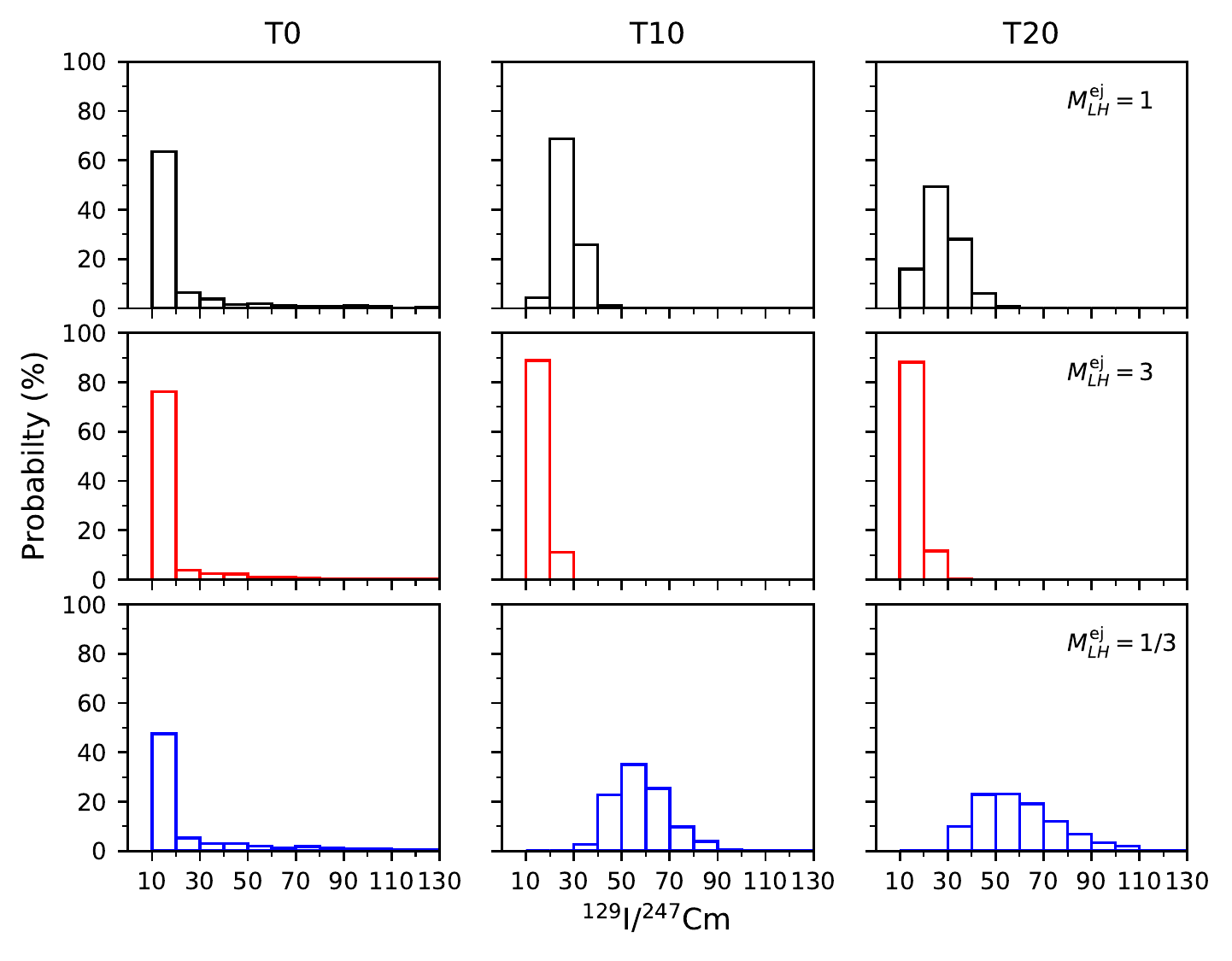}}
\caption{Probability distribution of $^{129}$I/$^{127}$Cm ratio in the ESS for two equally frequent \textsl{r}-process sources $S_L$ and $S_H$ with $\lambda_L=10$ and $\lambda_H=1000$ for  $M_{LM}^{\rm ej}=1$ (black), 3 (red), and 1/3 (blue) and tolerances T0 (left vertical panel), T10 (middle vertical panel), and T20 (right vertical panel) corresponding to the values listed in Table~\ref{tab:2sourcesLH}. All models have $D=0.1$ kpc$^2$ Gyr$^{-1}$.}

\label{fig:Dp1_LH_hist}
\end{figure*}

The dramatic change in the probability distribution when we apply the T10 or T20 criteria is primarily due to the requirement of concordance for $^{129}$I/$^{127}$I and $^{247}$Cm/$^{235}$U ratios. 
Because $^{129}$I and $^{247}$Cm have almost identical lifetimes and, $^{127}$I and $^{235}$U are stable or relatively long-lived, the relative ratio of $^{129}$I/$^{127}$I to $^{247}$Cm/$^{235}$U 
is essentially unaffected by the free decay during the typical isolation time  $\Delta_{\rm iso}\lesssim 50$ Myr. 
Therefore, the criteria for concordance requires the  $^{129}$I/$^{127}$I and $^{247}$Cm/$^{235}$U ratios to overlap within the tolerance level at $t_{\rm iso}$. In the scenario where $\nu_L=\nu_M$, both type of sources contribute equally to $^{127}$I whereas $^{235}$U is mostly from the $S_L$ source as it produces ten times more $^{235}$U per event relative to $S_M$. In addition, if we consider the simplified approximation where the SLRs $^{129}$I and $^{247}$Cm mostly come from one major event, the contribution of this event to the total amount of  $^{127}$I and $^{235}$U accumulated from all past events is negligible. 
Thus, the $^{129}$I/$^{127}$I ratio after the last major event is the same for both $S_M$ and $S_L$ as both sources produce identical amounts of $^{129}$I. 
However, the $^{247}$Cm/$^{235}$U ratio is a factor of $\sim 10$ lower if the last major event is $S_M$ compared to $S_L$ as the former produces 10 times less $^{247}$Cm.
With the adopted production factors, only the scenario where the last major event is $S_L$ can result 
in $^{129}$I/$^{127}$I ratio 
comparable to that of $^{247}$Cm/$^{235}$U at $(R_\odot,t_\odot)$.
If the last major event is $S_M$, the $^{247}$Cm/$^{235}$U is $\sim$ an order of magnitude lower, which substantially lowers the chances of concordance. 
This leads to  
a probability distribution of $^{129}$I/$^{247}$Cm ratio that is
mostly limited to low to intermediate values for criteria T10 and T20.  

The criteria for concordance also affects the fraction contributed from the last three highest contributing events. Table~\ref{tab:h1h2h3_concordance} shows the mean values of $f_{\rm h1},~f_{\rm h2},$ and $f_{\rm h3}$ for $^{129}$I with the concordance criteria T10 and T20 along with the the minimum criteria T0. Overall, compared to T0, for T10 and T20, the fraction contributed from the highest contributing event h1 decreases while the fraction contributed from h2 and h3 increases. With the concordance criteria, the mean values $f_{\rm h1},~f_{\rm h2},$ and $f_{\rm h3}$ ranges form $\sim 0.53$--0.76, 0.17--0.34, and 0.04--0.12 compared to $0.70$--0.84, 0.12--0.17, and 0.03--0.06, respectively, for the minimum criteria.

\begin{table*}
\caption{Probability distribution of $^{129}$I/$^{247}$Cm ratio in the ESS for concordance criteria T0, T10 and T20 (see text) for three equally frequent \textsl{r}-process sources: one MRSN like $S_H$ source with $\lambda_H=1000$, one BNSM like $S_M$ source with $\lambda_M=100$, and one BNSM like $S_L$ source with $\lambda_L=10$.
} 
\centering 
\begin{tabular}{c c c c c c c c c c c c} 
%\begin{tabular}{|p{2.0cm}|p{1.4cm}|p{.9cm}|p{.9cm}|p{.9cm}|p{1cm}|p{.9cm}|p{.9cm}|p{1cm}|p{1cm}|p{1cm}|p{.9cm}}
\hline\hline 
Model& &\multicolumn{10}{c|}{Probability of $^{129}$I/$^{247}$Cm within an interval }\\ [0.5ex]
\hline
$(D,M^{\rm ej}_{LH},M^{\rm ej}_{LM})$ &Criteria & 10-20 & 20-30 & 30-50 &50-70 & 70-80 & 80-90 & 90-110 &110-130&130-150& $>150$\\ [0.5ex] 
\hline 
\multirow{3}{*}{(0.1,1,1)}  &T0 & 0.44& 0.05 & 0.06 & 0.03 & 0.02 & 0.04 & 0.18 & 0.02 & 0.02&0.14\\
&T10 & 0.00 & 0.19 &0.76 &0.05 &0.00 &0.00 & 0.00 &0.00 &0.00&0.00\\
&T20 & 0.02 & 0.25 &0.61 &0.12 &0.01 &0.00 & 0.00 &0.00 &0.00&0.00\\
%&T40 & 0.16 &0.20 &0.23 &0.15 &0.07 &0.05 &0.14 &0.01 &0.00&0.00\\
\hline
\multirow{3}{*}{(0.3,1,1)}  &T0 & 0.40 &0.10& 0.11& 0.06& 0.03& 0.04& 0.10& 0.02& 0.01& 0.13 \\
&T10 & 0.00& 0.16& 0.84& 0.00& 0.00& 0.00& 0.00& 0.00& 0.00& 0.00  \\
&T20 & 0.00& 0.32& 0.61& 0.07& 0.00& 0.00& 0.00& 0.00& 0.00& 0.00  \\
%&T40 & 0.16& 0.23& 0.29& 0.21& 0.07& 0.02& 0.02& 0.00& 0.00& 0.00  \\
\hline
\hline
\multirow{3}{*}{(0.1,3,3)}  &T0 & 0.69 &0.04 &0.05 &0.03 &0.01 &0.02 &0.11 &0.01 & 0.01&0.05\\
&T10 &0.87 &0.13 & 0.00& 0.00& 0.00& 0.00& 0.00& 0.00& 0.00 &  0.00\\
&T20 &0.86 &0.14 & 0.00& 0.00& 0.00& 0.00& 0.00& 0.00& 0.00 &  0.00\\
%&T40 &0.83 &0.11 &0.07& 0.00& 0.00& 0.00& 0.00& 0.00& 0.00  & 0.00\\
\hline
\multirow{3}{*}{(0.3,3,3)}  &T0 & 0.6& 0.09& 0.08& 0.05& 0.03& 0.03& 0.05&  0.01& 0.00& 0.05\\
&T10 &0.91 &0.09 & 0.00& 0.00& 0.00& 0.00& 0.00& 0.00& 0.00 &  0.00\\
&T20 &0.84 &0.16 & 0.00& 0.00& 0.00& 0.00& 0.00& 0.00& 0.00 &  0.00\\
%&T40 &0.73 &0.18 & 0.08& 0.00& 0.00& 0.00& 0.00& 0.00& 0.00 &  0.00\\
\hline
\hline
\multirow{3}{*}{(0.1,1,1/3)}  &T0 & 0.31& 0.05&  0.04& 0.05&  0.03&  0.02& 0.22& 0.03& 0.01&0.23\\
&T10 & 0.00& 0.00 &0.01& 0.25 & 0.18& 0.16& 0.40& 0.01& 0.00 &0.00\\
&T20 & 0.00& 0.00 &0.03& 0.13 & 0.09& 0.08& 0.61& 0.05& 0.01 &0.00\\
%&T40 & 0.00& 0.01 &0.08& 0.09 & 0.05& 0.07& 0.52& 0.11& 0.05 &0.03\\
\hline
\multirow{3}{*}{(0.3,1,1/3)}  &T0 &0.21 & 0.08 &0.09& 0.07&  0.03& 0.04& 0.17& 0.04& 0.02& 0.26\\
&T10 & 0.00& 0.00 &0.00& 0.50&  0.31& 0.16& 0.04 &0.00& 0.00&  0.00 \\
&T20 & 0.00& 0.00 &0.06& 0.25&  0.14& 0.15& 0.38 &0.02& 0.00&  0.00 \\
%&T40 & 0.00& 0.01& 0.14& 0.14& 0.08& 0.08& 0.35& 0.12& 0.06& 0.02 \\
\hline
\hline
\multirow{3}{*}{(0.1,1/3,1)}  &T0& 0.36& 0.06& 0.06& 0.03& 0.02& 0.02& 0.16& 0.02& 0.01&0.25\\ 
&T10 & 0.00 &0.00& 0.09& 0.46 &0.15 &0.13 &0.17& 0.01& 0.00&0.00 \\
&T20 & 0.00 &0.00& 0.11& 0.20 &0.12 &0.12 &0.42& 0.04& 0.00&0.00 \\
%&T40 & 0.00& 0.04& 0.11& 0.10& 0.06& 0.05& 0.47& 0.10& 0.05&0.03\\
\hline
\multirow{3}{*}{(0.3,1/3,1)}  &T0&  0.28 &0.09& 0.10& 0.06& 0.03& 0.03 & 0.09 & 0.02& 0.03& 0.27\\
&T10 & 0.00 &0.00& 0.12& 0.75& 0.11& 0.02&  0.00& 0.00& 0.00& 0.00 \\
&T20 & 0.00 &0.00& 0.21& 0.39& 0.19& 0.12&  0.09& 0.00& 0.00& 0.00 \\
%&T40 & 0.00 &0.05& 0.20& 0.15& 0.08& 0.10& 0.27& 0.10& 0.03& 0.00  \\

\hline
\end{tabular}
\label{tab:3sources} 
\end{table*}

\subsection{Results with Two Sources of Type $S_L$ and $S_H$}

We also explored the scenario with two sources $S_L$ and $S_H$, where the former is a BNSM like source and the latter is a MRSN like source with $\lambda_L=10$ and $\lambda_H=1000$.
As before, we assume that both sources are equally frequent and consider three different relative ejecta mass ratios of $M_{LH}^{\rm ej}=1/3, 1,$ and 3. In the case of minimum criteria T0, the probability distribution has a peak at values close to $\lambda_L$ with a very long tail that extends up to $\lambda_H$. The reason is similar to the scenario discussed before with $S_L$ and $S_M$. Even if the main contributing h1 event is of $S_H$ type, a minor contribution from $S_L$ either as h2 or h3 produces a significant amount of $^{247}$Cm and decreases the value of the $^{129}$I/$^{247}$Cm ratio and brings it closer to $\lambda_L$. When the criteria for concordance i.e, T10 or T20 is imposed, the probability distribution for the  $^{129}$I/$^{247}$Cm again changes drastically (see Fig.~\ref{fig:Dp1_LH_hist}). In this case, the probability distribution is primarily limited to values  $\lesssim 7\lambda_L$  and is negligible for values $\gtrsim \lambda_M$ in all cases. When $M_{LH}^{ej}=3$, the distribution  
strongly peaks at $\sim \lambda_L$, whereas for $M_{LH}^{ej}=1$, the peak of the probability is at $\sim 2-5\lambda_L$. Even when $S_H$ is the dominant source with $M_{LH}^{ej}=1$, the peak of the probability distribution is at $\sim 5-7\lambda_L$ with zero probability for values $>11\lambda_L$. 

The reason for the dramatic change in the probability distribution when T10 and T20 is imposed is again similar to the scenario discussed before with $S_L$ and $S_M$. The allowed values for the $^{129}$I/$^{247}$Cm ratio is governed by the requirement of concordance for $^{129}$I/$^{127}$I and $^{247}$Cm/$^{235}$U. As before, both sources contribute equally to the total $^{127}$I but almost all of the $^{235}$U comes from $S_L$. Thus, if the last few events are dominated by $S_H$, the $^{247}$Cm/$^{235}$U ratio is factor $\sim 100$ lower than $^{129}$I/$^{127}$I which makes concordance impossible.

\subsection{Results with Two Sources of Type $S_M$ and $S_H$}
Because we are dealing with isotopic ratios, the results for the evolution of $^{129}$I/$^{247}$Cm ratio for a fixed value of $\lambda_i/\lambda_j$ and $\nu_i/\nu_j$ can be used to get the corresponding results for $\kappa\lambda_i$ and $\kappa\lambda_j$ by simply multiplying the results for $^{129}$I/$^{247}$Cm the corresponding factor $\kappa$. Thus, the results for the case of two types of sources $S_L$ and $S_M$ with $\lambda_L=10$ and $\lambda_M=100$ can be used to get the results for the case with two sources $S_M$ and $S_H$ with $\lambda_M=100$ and $\lambda_H=1000$ with $\nu_{MH}=1$ by scaling up the probability distribution of $^{129}$I/$^{247}$Cm ratio by a factor of 10. In this case, with the criteria T10 and T20, the probability distribution is mostly limited to values of $\lesssim 5\lambda_M$ and strongly peaks at values of $\sim 300$--500.

\begin{figure*}
\centerline{\includegraphics[width=\textwidth]{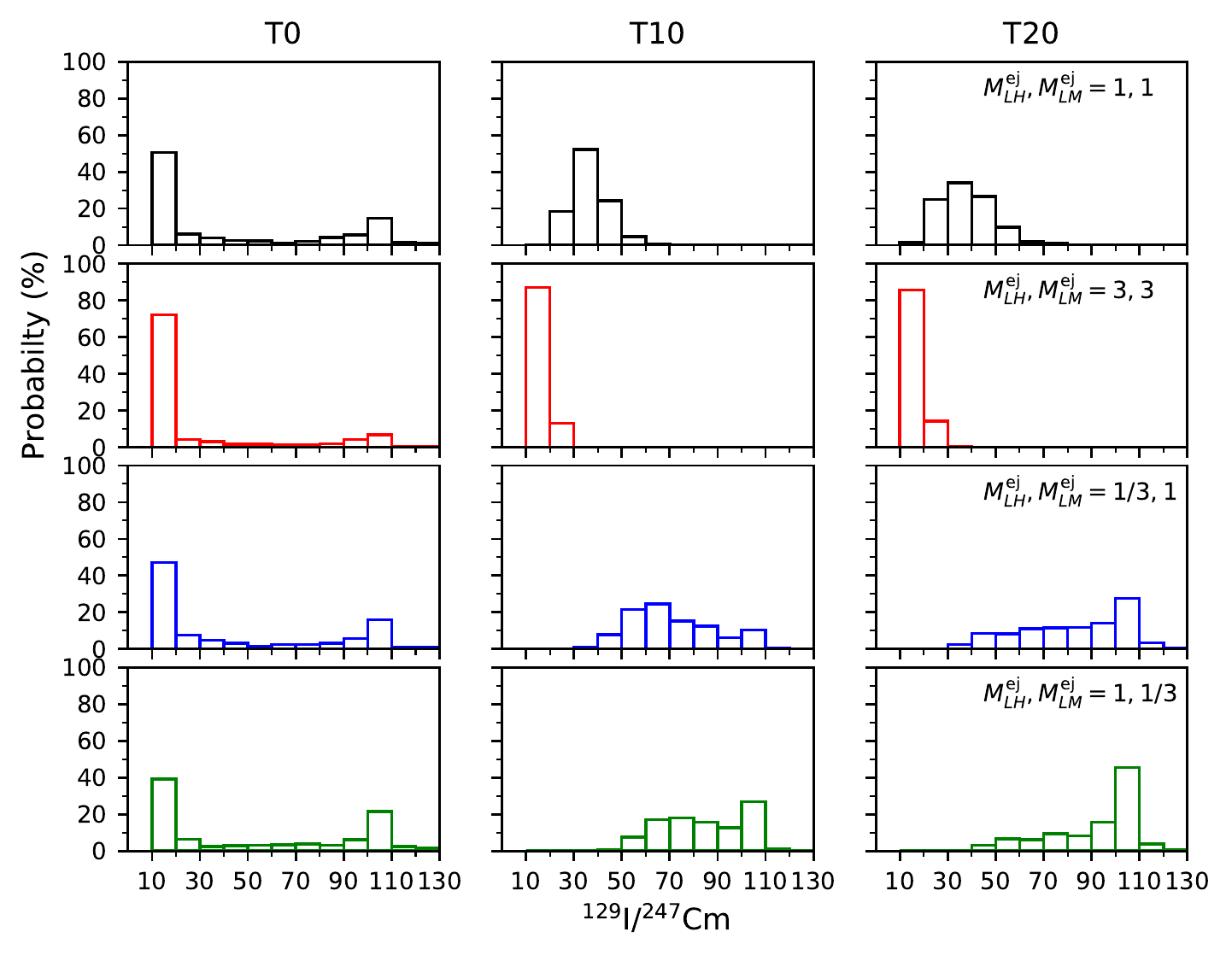}}
\caption{Probability distribution of $^{129}$I/$^{127}$Cm ratio in the ESS for three equally frequent \textsl{r}-process sources $S_L$, $S_M$, and $S_H$ with $\lambda_L=10$, $\lambda_M=100$, and $\lambda_H=1000$, respectively, for  $(M_{LM}^{\rm ej},M_{LH}^{\rm ej})=(1,1)$ (black), $(3,3)$ (red), $(1/3,1)$ (blue), and $(1,1/3)$ (green) and tolerances T0 (left vertical panel), T10 (middle vertical panel), and T20 (right vertical panel) corresponding to the values listed in Table~\ref{tab:3sources}. All models have $D=0.1$ kpc$^2$ Gyr$^{-1}$.
}

\label{fig:Dp1_LMH_hist}
\end{figure*}
\subsection{Result with Three Sources}
Finally, we consider the scenario with three different sources $S_L$, $S_M$, and $S_H$ with distinct values for $^{129}$I/$^{247}$Cm ratio covering two orders of magnitude ranging from low $\lambda_L=10$, to medium $\lambda_M=100$, to high $\lambda_H=1000$. 
The frequency for all sources are taken to be equal. In this case, there are two parameters for the ratio of mass of \textsl{r}-process ejecta. They are denoted by $M_{LH}^{\rm ej}$ and  $M_{LM}^{\rm ej}$ corresponding to the ejecta mass ratio for $S_L$ relative to $S_H$ and $S_L$ relative to $S_M$, respectively. We simulate four different scenarios. The first one is where all sources contribute equally to the main \textsl{r}-process, i.e, $M_{LH}^{\rm ej}=M_{LM}^{\rm ej}=1$. The other three scenarios have one of the sources as the dominant main \textsl{r}-process source, which has three times more ejecta mass than the rest of the two sources.
As before, for each scenario, we consider two different values of $D$. The probability distribution of the $^{129}$I/$^{247}$Cm ratio for the three different criteria T0, T10, and T20 are listed in Table~\ref{tab:3sources} along with the corresponding figure for $D=0.1$ kpc$^2$ Gyr$^{-1}$ in Fig.~\ref{fig:Dp1_LMH_hist}.

The overall results are roughly an average of the results derived for the scenario with two sources ($S_L$, $S_M$) and  ($S_L$, $S_H$).
When the minimum criteria T0 is applied, the probability distribution for the  $^{129}$I/$^{247}$Cm ratio has the highest peak close to $\lambda_L$ with values ranging from $\sim 0.2$--$0.7$, and a relatively smaller peak at $\sim \lambda_M$ with values ranging from $\sim 0.09$--$0.22$ along with probability for intermediate values between $\lambda_L$ and $\lambda_M$ ranging from $\sim 0.09$-$0.17$. The probability distribution drops sharply for values $\gtrsim \lambda_M$ and has a long and relative flat tail that extends to values up to $\sim \lambda_H$. 
When the criteria for concordance are imposed, the probability distribution changes drastically similar to the cases with two sources. Firstly, the long tail of the probability distribution above $\gtrsim \lambda_M$ vanishes completely. As with the models with two sources, this change in the probability distribution is due to the requirement of concordance for $^{129}$I/$^{127}$I and $^{247}$Cm/$^{235}$U such that if the last major contributors are events of the $S_H$ type, concordance is impossible. 

There is, however, a significant difference in the probability distribution for 3 sources when compared to results from 2 sources. In the case of 2 sources, for all models, the probability distribution peaks at values  $\sim \lambda_L=10$ or intermediate values of $\sim 50$ with negligible or very low  probability for values close to $\lambda_M$. In contrast, for 3 sources, the probability distribution has either peaks or relatively high values close to $\lambda_M$ for models where $S_M$ or $S_H$ are the dominant \textsl{r}-process source. However, in all such models, there is a substantial probability ranging from 0.16--0.60 for the ratio to have intermediate values between 30--70. In the model with equal ejecta mass for all sources, the probability distribution peaks at intermediate values. Only for the model where $S_L$ is the dominant \textsl{r}-process source, the entire probability distribution  is limited close to $\lambda_L$ with a negligible probability for values $\gtrsim 30$.

\subsection{Isolation Time and Time of ``Last" Event}
\begin{figure*}
\centerline{\includegraphics[width=\textwidth]{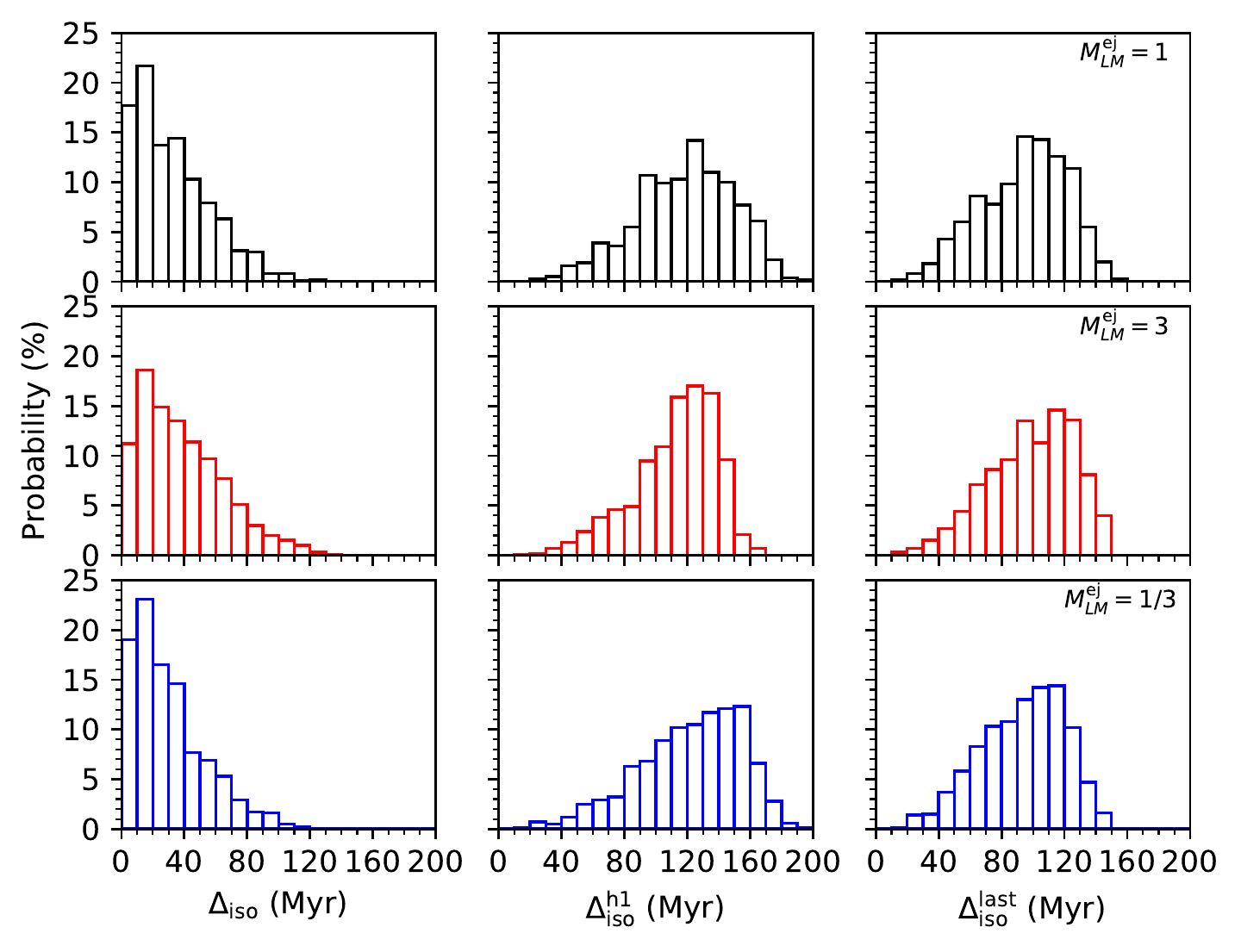}}
\caption{Probability distribution of $\Delta_{\rm iso}$, $\Delta_{\rm iso}^{\rm h1}$, and $\Delta_{\rm iso}^{\rm last}$ for two equally frequent \textsl{r}-process sources $S_L$ and $S_M$ with $\lambda_L=10$ and $\lambda_M=100$ for $M_{LM}^{\rm ej}=1$ (black), 3 (red), and 1/3 (blue). All models have $D=0.1$ kpc$^2$ Gyr$^{-1}$ and tolerance T20.}

\label{fig:Delta_LM_hist}
\end{figure*}

It is interesting to consider the typical isolation time $\Delta_{\rm iso}$ in our models. In addition, we can also consider the time interval $\delta_{\rm h1}$ between the highest $^{129}$I contributing event and the beginning of isolation at $t_{\rm iso}$. 
Among the three highest contributors, h1, h2, and h3, h1 is not necessarily the last contributing event. Thus we additionally define $\delta_{\rm last}$ as the time interval between the most recent event among h1, h2, and h3 and $t_{\rm iso}$. The effective isolation time corresponding to h1 and the last event can be defined as $\Delta_{\rm iso}^{\rm h1}= \Delta_{\rm iso}+\delta_{\rm h1}$ and $\Delta_{\rm iso}^{\rm last}= \Delta_{\rm iso}+\delta_{\rm last}$. We find that the distribution of $\Delta_{\rm iso}$, $\Delta_{\rm iso}^{\rm h1}$, and $\Delta_{\rm iso}^{\rm last}$ are similar for the four different scenarios consider in this work. For the purpose of illustration, the distribution of $\Delta_{\rm iso}$, $\Delta_{\rm iso}^{\rm h1}$, and $\Delta_{\rm iso}^{\rm last}$ are shown in Fig.~\ref{fig:Delta_LM_hist} for the scenario with two sources $S_L$ and $S_M$ for models with $D=0.1$ kpc$^2$ Gyr$^{-1}$ and criteria T20. 
The distribution of $\Delta_{\rm iso}$ peaks at values of $\sim \lesssim 50$~Myr which is consistent with the typical lifetimes associated with molecular and giant molecular clouds \citep{hartmann2001,murray2011}. The 68th percentile range for $\Delta_{\rm iso}^{\rm h1}$ ranges form $\sim 100$--140 Myr whereas the corresponding range for $\Delta_{\rm iso}^{\rm last}$ is $\sim 80$--115 Myr. The central values are broadly consistent with recent calculations by \citet{cote2019} where the total isolation time for $^{129}$I was found to be $\sim 85$--116 Myr. Interestingly, in our case, the probability for the effective isolation time to be $\lesssim 50$ Myr is low with typical values $\lesssim 5\%$ but not zero.    
 
\section{Discussion \& Conclusions}

In this work we explored the evolution of the SLRs produced by \textsl{r}-process at the solar location due to two or three different \textsl{r}-process sources that have distinct $^{129}$I/$^{247}$Cm production ratios. In contrast to the conclusion reached in \citet{Cote2021}, we find that in general, the observed ESS ratio for $^{129}$I/$^{247}$Cm does not correspond to a single ``last'' event. Although there is a major contributing last event, it accounts for only $\sim 50$--75\% of all the $^{129}$I in the ESS and at least two more minor contributing events are required to account for $\gtrsim 95\%$ of the observed $^{129}$I. This has a large impact on the probability distribution for the $^{129}$I/$^{247}$Cm ratio in the ESS
that depends on the particular choice of parameters such as the ratio of ejecta masses, and the relative frequency of the sources. 

One of the reasons for the difference in our conclusion and the one reached by \citet{Cote2021} is 
related to the
prescription for modelling the evolution of \textsl{r}-process elements. 
The turbulent gas diffusion prescription used in this work is very different from the one used in \citet{Cote2021} where a stochastic one-zone model was used. Although both prescriptions are able to model the stochasticity of occurrence time of \textsl{r}-process events, the diffusion prescription can also capture the stochasticity associated with the spatial location, distance of the event from the solar location as well as the corresponding dilution. The other important reason is the application of the criteria for concordance used in our work, which %that
involves using the ESS ratio of $^{129}$I/$^{127}$I and $^{247}$Cm/$^{235}$U that is different from what was considered in \citet{Cote2021}. 
As mentioned earlier, the concordance imposes additional constraints that changes the ESS $^{129}$I/$^{247}$Cm ratio substantially..

Although we find that the ESS $^{129}$I/$^{247}$Cm ratio measured in meteorites cannot be used to directly constrain the ``last'' \textsl{r}-process event, the concordance criteria used in our work still allows us to put interesting constraints on the nature of \textsl{r}-process sources when combined with theoretical nucleosynthetic calculations for various astrophysical sources. 
Below, we take the results of \textsl{r}-process nucleosynthesis calculations for the various astrophysical scenarios and nuclear physics inputs by \citet{Cote2021} as examples for illustration and discussion. 
The astrophysical models considered in \citet{Cote2021} are, i) dynamical ejecta from NS-NS and NS-BH mergers that are the most neutron rich and have the lowest values of $^{129}$I/$^{247}$Cm, ii) three different NS-NS merger disk ejecta numbered 1, 2 and 3 with varying levels of neutron richness resulting in higher values of $^{129}$I/$^{247}$Cm compared to the dynamical ejecta, and iii) MRSN ejecta that have the highest values of $^{129}$I/$^{247}$Cm. The values of $^{129}$I/$^{247}$Cm in all models are sensitive to the nuclear physics inputs where three different nuclear reaction rates and three different fission fragment distributions were considered for each model. 
The ranges of the $^{129}$I/$^{247}$Cm ratio in these models are roughly between 10--100 for the neutron-rich dynamical ejecta, 250--1000 for disk ejecta 1, 50--250 for disk ejecta 2, 25--150 for disk ejecta 3, and 1000-6000 for MRSN ejecta.
We associate  
the dynamical ejecta from NS-NS and NS-BH mergers as the $S_L$ source, one of the three different disk ejecta as the $S_M$ source, and the MRSN ejecta as the $S_H$ source.

Considering two equally frequent \textsl{r}-process sources $S_L$ and $S_M$ with 
$\lambda_M/\lambda_{L}\approx 10$, 
we can draw the following conclusions when the concordance criteria (T10 or T20) is imposed;
\begin{itemize}
\item Because the probability distribution can have a maximum value of $\sim \lambda_M$, the possibility of NS-NS merger disk ejecta 3 as $S_M$ is ruled out. 
This is simply due to the fact that for all NS-NS merger disk ejecta 3 models, $\lambda \lesssim 150$ making it incompatible with the observed value of $438\pm 184$.

\item The $S_L$ source as the NS-NS or NS-BH dynamical ejecta being the dominant \textsl{r}-process source (i.e, $M_{LM}^{\rm ej}=3$) is highly disfavoured for all models. This is because in this case the values for the $^{129}$I/$^{247}$Cm ratio is limited to $< 3\lambda_L$ whereas the $^{129}$I/$^{247}$Cm production ratio $\lesssim 90$ for all dynamical ejecta models considered by \citep{Cote2021}\footnote{Other than the TF(D3C*) model which is an outlier.}. This is inconsistent with the measured value of $438\pm 184$. 

\item If the $S_M$ source is the dominant contributor to the \textsl{r}-process (i.e, $M_{LM}^{\rm ej}=1/3$),  NS-NS merger disk ejecta 1 is favoured while NS-NS merger disk ejecta 2 is marginally consistent with the observed values.

\item  The $S_L$ source as NS-NS or NS-BH dynamical ejecta as an equal contributor to the \textsl{r}-process (i.e., $M_{LM}^{\rm ej}=1$) requires the $S_M$ source to be  NS-NS merger disk ejecta 1. This is because when $S_L$ is an equal contributor, the probability distribution  for the $^{129}$I/$^{247}$Cm ratio is limited to values of $\lesssim \lambda_M/2$. Among all the disk models, this can only be satisfied by  the NS-NS merger disk ejecta 1 where the $^{129}$I/$^{247}$Cm production ratio can reach values of up to $\sim 900$ such that the probability distribution for the $^{129}$I/$^{247}$Cm ratio is non-negligible at the observed value of $438\pm 184$. 

\item Overall, the most favoured scenario is the NS-NS merger disk ejecta 1 as the dominant $S_M$ source which gives the highest probability for the  $^{129}$I/$^{247}$Cm ratio to be in the range consistent with the observed value of $438\pm 184$.

\end{itemize}

For the case with two equally frequent \textsl{r}-process sources $S_M$ and $S_H$ with $\lambda_H/\lambda_M\approx 10$ and MRSN ejecta the $S_H$ source, all the three NS-NS merger disk ejecta  are possible as they result in probability distribution of $^{129}$I/$^{247}$Cm that is compatible with the observed value of $438\pm 184$. 

On the other hand, with MRSN ejecta the $S_H$ source and NS-NS or NS-BH dynamical ejecta as the $S_L$ source with $\lambda_H/\lambda_L\approx 100$ we conclude that;

\begin{itemize}
    \item NS-NS or NS-BH dynamical ejecta ($S_L$) as the dominant \textsl{r}-process source is highly disfavoured. Again, in this case the $^{129}$I/$^{247}$Cm ratio is limited to $\lesssim 3\lambda_L$ whereas $^{129}$I/$^{247}$Cm production ratio $\lesssim 90$ for all dynamical ejecta models considered by \citep{Cote2021}. This makes it incompatible with the observed value of $438\pm 184$.
    \item  NS-NS or NS-BH dynamical ejecta ($S_L$) as either equal or subdominant contributor to total \textsl{r}-process is consistent with the observed value.
    
\end{itemize}

Finally, with three equally frequent \textsl{r}-process sources with NS-NS or NS-BH dynamical ejecta as $S_L$, NS-NS merger disk ejecta as $S_M$ and MRSN ejecta as $S_H$, the conclusions we draw are;
\begin{itemize}

\item The probability distribution for the $^{129}$I/$^{247}$Cm ratio is limited to values between  $\lambda_L$ to $\sim \lambda_M$. Thus, the ESS value of $438\pm 184$ disfavours NS-NS merger disk ejecta 3 and and is marginally consistent with NS-NS merger disk ejecta 2 as the $S_M$ source whereas NS-NS merger disk ejecta 1 is favoured.

\item Similar to the scenario with two sources, the $S_L$ source as the NS-NS or NS-BH dynamical ejecta being the dominant \textsl{r}-process source (i.e $M_{LH}^{\rm ej},M_{LM}^{\rm ej}=3$) is highly disfavoured for all models. 

\item Overall, the most favoured scenario involves NS-NS merger disk ejecta 1 as the $S_M$ source,  with either the $S_H$ (MRSN ejecta) or the disk ejecta are the dominant \textsl{r}-process source (i.e, $M_{LH}^{\rm ej},M_{LM}^{\rm ej}=1/3,1$ or $M_{LH}^{\rm ej},M_{LM}^{\rm ej}=1,1/3$).  
\end{itemize}

\section{Summary and Outlook}

We studied the evolution of \textsl{r}-process isotopes including SLRs at the birth location of the sun and the prospect of using the $^{129}$I/$^{247}$Cm ratio to constrain the \textsl{r}-process sources. We find that the measured meteoritic value of the $^{129}$I/$^{247}$Cm ratio does not correspond to a single ``last'' \textsl{r}-process event when there are multiple sources with distinct $^{129}$I/$^{247}$Cm production ratios.
Instead, we find that the $^{129}$I/$^{247}$Cm ratio can be used to put important constraints on \textsl{r}-process sources when the ESS data of  $^{129}$I/$^{127}$I and $^{247}$Cm/$^{235}$U is taken into account. In particular, based on the nucleosynthesis calculation by \citet{Cote2021} for various astrophysical sites, we find that  models of NS-NS or NS-BH dynamical ejecta that are neutron rich and have low $^{129}$I/$^{247}$Cm ratio cannot be the dominant source of \textsl{r}-process. This statement holds both in the case of where there are just two sources as well as when all three sources are considered. 
Interestingly, this is consistent with current detailed BNSM merger simulations which predict that merger disk ejecta to be a more dominant source of \textsl{r}-process than the dynamical ejecta~\citep{Shibata:2019wef,Metzger:2019zeh}. 
If there is a MRSN like source that has a high value for the production ratio of $^{129}$I/$^{127}$I and is as frequent as BNSMs, then it can mix with the dynamical ejecta to produce values that are compatible with observations without the need for disk ejecta. 
However, from theoretical expectations, this is unlikely as substantial amount of \textsl{r}-process is expected from NS-NS merger disk ejecta. 
In a realistic scenario where the disk ejecta is as frequent as the dynamical ejecta, NS-NS merger disk ejecta  
which has medium values of $^{129}$I/$^{247}$Cm production ratio is highly favoured when there are no MRSN like source that is similarly  
frequent. However, if there is a MRSN like source, any of the current merger disk models calculated in \citet{Cote2021} are in fact possible.

Our analysis is based on simplifying assumptions such as equal frequency for all sources and fixed values of $\lambda$ for the three types of \textsl{r}-process sources and limited values of ejecta masses. In principle, future studies can explore a larger parameter space to identify allowed regions that are consistent with the ESS data.
However, for some realistic scenarios which have parameters that are not directly covered in this work, the results presented here can be easily extrapolated. For example, in the scenario with all the three sources, if the frequency of MRSN like event is lower, then it would effectively reduce to the scenario with two sources of type $S_L$ and $S_M$ and the conclusion drawn for such a scenario can be applied in this case. Similarly, if there are two different $S_M$ sources that have similar values of $\lambda$, then they can effectively be treated as a single $S_M$ source and the results for two sources with $S_L$ and $S_M$ can be applied in this case.  
In future, with better simulations of astrophysical sites with improved nuclear physics inputs and accurate nucleosynthetic models, the ESS ratio of $^{129}$I/$^{247}$Cm along with $^{129}$I/$^{127}$I and $^{247}$Cm/$^{235}$U could be used to provide strong constraint that could help identify the main source of \textsl{r}-process and even rule out certain astrophysical sites.

\section*{Acknowledgements}
M.-R.~W. acknowledges supports from the Ministry of Science and Technology, Taiwan under Grant No.~110-2112-M-001 -050, the Academia Sinica under Project No.~AS-CDA-109-M11, and the Physics Division, National Center for Theoretical Sciences, Taiwan.

%The Acknowledgements section is not numbered. Here you can thank helpful colleagues, acknowledge funding agencies, telescopes and facilities used etc. Try to keep it short.

%%%%%%%%%%%%%%%%%%%%%%%%%%%%%%%%%%%%%%%%%%%%%%%%%%
\section*{Data Availability}
Data is available upon request.



%\bibliographystyle{mnras}
%\bibliography{reference} % if your bibtex file is called example.bib

\begin{thebibliography}{}
\makeatletter
\relax
\def\mn@urlcharsother{\let\do\@makeother \do\$\do\&\do\#\do\^\do\_\do\%\do\~}
\def\mn@doi{\begingroup\mn@urlcharsother \@ifnextchar [ {\mn@doi@}
  {\mn@doi@[]}}
\def\mn@doi@[#1]#2{\def\@tempa{#1}\ifx\@tempa\@empty \href
  {http://dx.doi.org/#2} {doi:#2}\else \href {http://dx.doi.org/#2} {#1}\fi
  \endgroup}
\def\mn@eprint#1#2{\mn@eprint@#1:#2::\@nil}
\def\mn@eprint@arXiv#1{\href {http://arxiv.org/abs/#1} {{\tt arXiv:#1}}}
\def\mn@eprint@dblp#1{\href {http://dblp.uni-trier.de/rec/bibtex/#1.xml}
  {dblp:#1}}
\def\mn@eprint@#1:#2:#3:#4\@nil{\def\@tempa {#1}\def\@tempb {#2}\def\@tempc
  {#3}\ifx \@tempc \@empty \let \@tempc \@tempb \let \@tempb \@tempa \fi \ifx
  \@tempb \@empty \def\@tempb {arXiv}\fi \@ifundefined
  {mn@eprint@\@tempb}{\@tempb:\@tempc}{\expandafter \expandafter \csname
  mn@eprint@\@tempb\endcsname \expandafter{\@tempc}}}

\bibitem[\protect\citeauthoryear{{Abbott} et~al.,}{{Abbott}
  et~al.}{2017a}]{abbott2017a}
{Abbott} B.~P.,  et~al., 2017a, \mn@doi [\prl]
  {10.1103/PhysRevLett.119.161101}, \href
  {https://ui.adsabs.harvard.edu/abs/2017PhRvL.119p1101A} {119, 161101}

\bibitem[\protect\citeauthoryear{{Abbott} et~al.,}{{Abbott}
  et~al.}{2017b}]{abbott2017b}
{Abbott} B.~P.,  et~al., 2017b, \mn@doi [\apjl] {10.3847/2041-8213/aa91c9},
  \href {https://ui.adsabs.harvard.edu/abs/2017ApJ...848L..12A} {848, L12}

\bibitem[\protect\citeauthoryear{{Arcones} \& {Thielemann}}{{Arcones} \&
  {Thielemann}}{2013}]{Arcones2013}
{Arcones} A.,  {Thielemann} F.~K.,  2013, \mn@doi [Journal of Physics G Nuclear
  Physics] {10.1088/0954-3899/40/1/013201}, \href
  {https://ui.adsabs.harvard.edu/abs/2013JPhG...40a3201A} {40, 013201}

\bibitem[\protect\citeauthoryear{{Banerjee}, {Wu}  \& {Yuan}}{{Banerjee}
  et~al.}{2020}]{bwy2020}
{Banerjee} P.,  {Wu} M.-R.,   {Yuan} Z.,  2020, \mn@doi [\apjl]
  {10.3847/2041-8213/abbc0d}, \href
  {https://ui.adsabs.harvard.edu/abs/2020ApJ...902L..34B} {902, L34}

\bibitem[\protect\citeauthoryear{Bartos \& Marka}{Bartos \&
  Marka}{2019}]{Bartos:2019cec}
Bartos I.,  Marka S.,  2019, \mn@doi [Nature] {10.1038/s41586-019-1113-7}, 569,
  85

\bibitem[\protect\citeauthoryear{{Beniamini} \& {Hotokezaka}}{{Beniamini} \&
  {Hotokezaka}}{2020}]{beniamini2020}
{Beniamini} P.,  {Hotokezaka} K.,  2020, \mn@doi [\mnras]
  {10.1093/mnras/staa1690}, \href
  {https://ui.adsabs.harvard.edu/abs/2020MNRAS.496.1891B} {496, 1891}

\bibitem[\protect\citeauthoryear{{Bovard}, {Martin}, {Guercilena}, {Arcones},
  {Rezzolla}  \& {Korobkin}}{{Bovard} et~al.}{2017}]{Bovard2017}
{Bovard} L.,  {Martin} D.,  {Guercilena} F.,  {Arcones} A.,  {Rezzolla} L.,
  {Korobkin} O.,  2017, \mn@doi [\prd] {10.1103/PhysRevD.96.124005}, \href
  {https://ui.adsabs.harvard.edu/abs/2017PhRvD..96l4005B} {96, 124005}

\bibitem[\protect\citeauthoryear{{Bovy}}{{Bovy}}{2015}]{galpy}
{Bovy} J.,  2015, \mn@doi [\apjs] {10.1088/0067-0049/216/2/29}, \href
  {https://ui.adsabs.harvard.edu/abs/2015ApJS..216...29B} {216, 29}

\bibitem[\protect\citeauthoryear{{Chiappini}, {Matteucci}  \&
  {Gratton}}{{Chiappini} et~al.}{1997}]{chiappini1997}
{Chiappini} C.,  {Matteucci} F.,   {Gratton} R.,  1997, \mn@doi [\apj]
  {10.1086/303726}, \href
  {https://ui.adsabs.harvard.edu/abs/1997ApJ...477..765C} {477, 765}

\bibitem[\protect\citeauthoryear{{C{\^o}t{\'e}} et~al.,}{{C{\^o}t{\'e}}
  et~al.}{2019a}]{cote2019a}
{C{\^o}t{\'e}} B.,  et~al., 2019a, \mn@doi [\apj] {10.3847/1538-4357/ab10db},
  \href {https://ui.adsabs.harvard.edu/abs/2019ApJ...875..106C} {875, 106}

\bibitem[\protect\citeauthoryear{{C{\^o}t{\'e}}, {Lugaro}, {Reifarth},
  {Pignatari}, {Vil{\'a}gos}, {Yag{\"u}e}  \& {Gibson}}{{C{\^o}t{\'e}}
  et~al.}{2019b}]{cote2019}
{C{\^o}t{\'e}} B.,  {Lugaro} M.,  {Reifarth} R.,  {Pignatari} M.,
  {Vil{\'a}gos} B.,  {Yag{\"u}e} A.,   {Gibson} B.~K.,  2019b, \mn@doi [\apj]
  {10.3847/1538-4357/ab21d1}, \href
  {https://ui.adsabs.harvard.edu/abs/2019ApJ...878..156C} {878, 156}

\bibitem[\protect\citeauthoryear{{C{\^o}t{\'e}} et~al.,}{{C{\^o}t{\'e}}
  et~al.}{2021}]{Cote2021}
{C{\^o}t{\'e}} B.,  et~al., 2021, \mn@doi [Science] {10.1126/science.aba1111},
  \href {https://ui.adsabs.harvard.edu/abs/2021Sci...371..945C} {371, 945}

\bibitem[\protect\citeauthoryear{{Cowan}, {Sneden}, {Lawler}, {Aprahamian},
  {Wiescher}, {Langanke}, {Mart{\'\i}nez-Pinedo}  \& {Thielemann}}{{Cowan}
  et~al.}{2021}]{Cowan2019}
{Cowan} J.~J.,  {Sneden} C.,  {Lawler} J.~E.,  {Aprahamian} A.,  {Wiescher} M.,
   {Langanke} K.,  {Mart{\'\i}nez-Pinedo} G.,   {Thielemann} F.-K.,  2021,
  \mn@doi [Reviews of Modern Physics] {10.1103/RevModPhys.93.015002}, \href
  {https://ui.adsabs.harvard.edu/abs/2021RvMP...93a5002C} {93, 015002}

\bibitem[\protect\citeauthoryear{{Cowperthwaite} et~al.,}{{Cowperthwaite}
  et~al.}{2017}]{Cowperthwaite+17}
{Cowperthwaite} P.~S.,  et~al., 2017, \mn@doi [\apjl]
  {10.3847/2041-8213/aa8fc7}, \href
  {https://ui.adsabs.harvard.edu/abs/2017ApJ...848L..17C} {848, L17}

\bibitem[\protect\citeauthoryear{{Fern{\'a}ndez} \& {Metzger}}{{Fern{\'a}ndez}
  \& {Metzger}}{2013}]{Fernandez2013}
{Fern{\'a}ndez} R.,  {Metzger} B.~D.,  2013, \mn@doi [\mnras]
  {10.1093/mnras/stt1312}, \href
  {https://ui.adsabs.harvard.edu/abs/2013MNRAS.435..502F} {435, 502}

\bibitem[\protect\citeauthoryear{{Fischer}, {Wu}, {Wehmeyer}, {Bastian},
  {Mart{\'\i}nez-Pinedo}  \& {Thielemann}}{{Fischer}
  et~al.}{2020}]{FischerWu2020}
{Fischer} T.,  {Wu} M.-R.,  {Wehmeyer} B.,  {Bastian} N.-U.~F.,
  {Mart{\'\i}nez-Pinedo} G.,   {Thielemann} F.-K.,  2020, \mn@doi [\apj]
  {10.3847/1538-4357/ab86b0}, \href
  {https://ui.adsabs.harvard.edu/abs/2020ApJ...894....9F} {894, 9}

\bibitem[\protect\citeauthoryear{{Foucart} et~al.,}{{Foucart}
  et~al.}{2014}]{Foucart2014}
{Foucart} F.,  et~al., 2014, \mn@doi [\prd] {10.1103/PhysRevD.90.024026}, \href
  {https://ui.adsabs.harvard.edu/abs/2014PhRvD..90b4026F} {90, 024026}

\bibitem[\protect\citeauthoryear{{Freiburghaus}, {Rosswog}  \&
  {Thielemann}}{{Freiburghaus} et~al.}{1999}]{Freiburghaus1999}
{Freiburghaus} C.,  {Rosswog} S.,   {Thielemann} F.~K.,  1999, \mn@doi [\apjl]
  {10.1086/312343}, \href
  {https://ui.adsabs.harvard.edu/abs/1999ApJ...525L.121F} {525, L121}

\bibitem[\protect\citeauthoryear{{Fujibayashi}, {Kiuchi}, {Nishimura},
  {Sekiguchi}  \& {Shibata}}{{Fujibayashi} et~al.}{2018}]{Fujibayashi2018}
{Fujibayashi} S.,  {Kiuchi} K.,  {Nishimura} N.,  {Sekiguchi} Y.,   {Shibata}
  M.,  2018, \mn@doi [\apj] {10.3847/1538-4357/aabafd}, \href
  {https://ui.adsabs.harvard.edu/abs/2018ApJ...860...64F} {860, 64}

\bibitem[\protect\citeauthoryear{{George}, {Wu}, {Tamborra}, {Ardevol-Pulpillo}
   \& {Janka}}{{George} et~al.}{2020}]{George2020}
{George} M.,  {Wu} M.-R.,  {Tamborra} I.,  {Ardevol-Pulpillo} R.,   {Janka}
  H.-T.,  2020, \mn@doi [\prd] {10.1103/PhysRevD.102.103015}, \href
  {https://ui.adsabs.harvard.edu/abs/2020PhRvD.102j3015G} {102, 103015}

\bibitem[\protect\citeauthoryear{{Goriely}, {Bauswein}  \& {Janka}}{{Goriely}
  et~al.}{2011}]{Goriely2011}
{Goriely} S.,  {Bauswein} A.,   {Janka} H.-T.,  2011, \mn@doi [\apjl]
  {10.1088/2041-8205/738/2/L32}, \href
  {https://ui.adsabs.harvard.edu/abs/2011ApJ...738L..32G} {738, L32}

\bibitem[\protect\citeauthoryear{{Grichener} \& {Soker}}{{Grichener} \&
  {Soker}}{2019}]{Aldana2019}
{Grichener} A.,  {Soker} N.,  2019, \mn@doi [\apj] {10.3847/1538-4357/ab1d5d},
  \href {https://ui.adsabs.harvard.edu/abs/2019ApJ...878...24G} {878, 24}

\bibitem[\protect\citeauthoryear{{Hartmann}, {Ballesteros-Paredes}  \&
  {Bergin}}{{Hartmann} et~al.}{2001}]{hartmann2001}
{Hartmann} L.,  {Ballesteros-Paredes} J.,   {Bergin} E.~A.,  2001, \mn@doi
  [\apj] {10.1086/323863}, \href
  {https://ui.adsabs.harvard.edu/abs/2001ApJ...562..852H} {562, 852}

\bibitem[\protect\citeauthoryear{Hotokezaka, Piran  \& Paul}{Hotokezaka
  et~al.}{2015a}]{Hotokezaka:2015zea}
Hotokezaka K.,  Piran T.,   Paul M.,  2015a, \mn@doi [Nature Phys.]
  {10.1038/nphys3574}, 11, 1042

\bibitem[\protect\citeauthoryear{{Hotokezaka}, {Piran}  \& {Paul}}{{Hotokezaka}
  et~al.}{2015b}]{hotokezaka2015}
{Hotokezaka} K.,  {Piran} T.,   {Paul} M.,  2015b, \mn@doi [Nature Physics]
  {10.1038/nphys3574}, \href
  {https://ui.adsabs.harvard.edu/abs/2015NatPh..11.1042H} {11, 1042}

\bibitem[\protect\citeauthoryear{{Hudson}, {Kennedy}, {Podosek}  \&
  {Hohenberg}}{{Hudson} et~al.}{1989}]{Hudson1989}
{Hudson} G.~B.,  {Kennedy} B.~M.,  {Podosek} F.~A.,   {Hohenberg} C.~M.,  1989,
  Lunar and Planetary Science Conference Proceedings, \href
  {https://ui.adsabs.harvard.edu/abs/1989LPSC...19..547H} {19, 547}

\bibitem[\protect\citeauthoryear{Just, Bauswein, Pulpillo, Goriely  \&
  Janka}{Just et~al.}{2015}]{Just:2014fka}
Just O.,  Bauswein A.,  Pulpillo R.~A.,  Goriely S.,   Janka H.~T.,  2015,
  \mn@doi [\mnras] {10.1093/mnras/stv009}, 448, 541

\bibitem[\protect\citeauthoryear{{Karakas}}{{Karakas}}{2010}]{karakas2010}
{Karakas} A.~I.,  2010, \mn@doi [\mnras] {10.1111/j.1365-2966.2009.16198.x},
  \href {https://ui.adsabs.harvard.edu/abs/2010MNRAS.403.1413K} {403, 1413}

\bibitem[\protect\citeauthoryear{{Kasen}, {Metzger}, {Barnes}, {Quataert}  \&
  {Ramirez-Ruiz}}{{Kasen} et~al.}{2017}]{Kasen+17}
{Kasen} D.,  {Metzger} B.,  {Barnes} J.,  {Quataert} E.,   {Ramirez-Ruiz} E.,
  2017, \mn@doi [\nat] {10.1038/nature24453}, \href
  {http://adsabs.harvard.edu/abs/2017Natur.551...80K} {551, 80}

\bibitem[\protect\citeauthoryear{{Kobayashi}, {Umeda}, {Nomoto}, {Tominaga}  \&
  {Ohkubo}}{{Kobayashi} et~al.}{2006}]{kobayashi2006}
{Kobayashi} C.,  {Umeda} H.,  {Nomoto} K.,  {Tominaga} N.,   {Ohkubo} T.,
  2006, \mn@doi [\apj] {10.1086/508914}, \href
  {https://ui.adsabs.harvard.edu/abs/2006ApJ...653.1145K} {653, 1145}

\bibitem[\protect\citeauthoryear{{Korobkin}, {Rosswog}, {Arcones}  \&
  {Winteler}}{{Korobkin} et~al.}{2012}]{Korobkin2012}
{Korobkin} O.,  {Rosswog} S.,  {Arcones} A.,   {Winteler} C.,  2012, \mn@doi
  [\mnras] {10.1111/j.1365-2966.2012.21859.x}, \href
  {https://ui.adsabs.harvard.edu/abs/2012MNRAS.426.1940K} {426, 1940}

\bibitem[\protect\citeauthoryear{Kullmann, Goriely, Just, Ardevol-Pulpillo,
  Bauswein  \& Janka}{Kullmann et~al.}{2021}]{Kullmann:2021gvo}
Kullmann I.,  Goriely S.,  Just O.,  Ardevol-Pulpillo R.,  Bauswein A.,   Janka
  H.~T.,  2021

\bibitem[\protect\citeauthoryear{{Kyutoku}, {Ioka}  \& {Shibata}}{{Kyutoku}
  et~al.}{2013}]{Kyutoku2013}
{Kyutoku} K.,  {Ioka} K.,   {Shibata} M.,  2013, \mn@doi [\prd]
  {10.1103/PhysRevD.88.041503}, \href
  {https://ui.adsabs.harvard.edu/abs/2013PhRvD..88d1503K} {88, 041503}

\bibitem[\protect\citeauthoryear{{Lattimer} \& {Schramm}}{{Lattimer} \&
  {Schramm}}{1974}]{Lattimer1974}
{Lattimer} J.~M.,  {Schramm} D.~N.,  1974, \mn@doi [\apjl] {10.1086/181612},
  \href {https://ui.adsabs.harvard.edu/abs/1974ApJ...192L.145L} {192, L145}

\bibitem[\protect\citeauthoryear{{Lugaro}, {Ott}  \& {Kereszturi}}{{Lugaro}
  et~al.}{2018}]{lugaro2018}
{Lugaro} M.,  {Ott} U.,   {Kereszturi} {\'A}.,  2018, \mn@doi [Progress in
  Particle and Nuclear Physics] {10.1016/j.ppnp.2018.05.002}, \href
  {https://ui.adsabs.harvard.edu/abs/2018PrPNP.102....1L} {102, 1}

\bibitem[\protect\citeauthoryear{{McKee}, {Parravano}  \& {Hollenbach}}{{McKee}
  et~al.}{2015}]{mckee2015}
{McKee} C.~F.,  {Parravano} A.,   {Hollenbach} D.~J.,  2015, \mn@doi [\apj]
  {10.1088/0004-637X/814/1/13}, \href
  {https://ui.adsabs.harvard.edu/abs/2015ApJ...814...13M} {814, 13}

\bibitem[\protect\citeauthoryear{{Mendoza-Temis}, {Wu}, {Langanke},
  {Mart{\'\i}nez-Pinedo}, {Bauswein}  \& {Janka}}{{Mendoza-Temis}
  et~al.}{2015}]{Mendoza2015}
{Mendoza-Temis} J. d.~J.,  {Wu} M.-R.,  {Langanke} K.,  {Mart{\'\i}nez-Pinedo}
  G.,  {Bauswein} A.,   {Janka} H.-T.,  2015, \mn@doi [\prc]
  {10.1103/PhysRevC.92.055805}, \href
  {https://ui.adsabs.harvard.edu/abs/2015PhRvC..92e5805M} {92, 055805}

\bibitem[\protect\citeauthoryear{Metzger}{Metzger}{2020}]{Metzger:2019zeh}
Metzger B.~D.,  2020, \mn@doi [Living Rev. Rel.] {10.1007/s41114-019-0024-0},
  23, 1

\bibitem[\protect\citeauthoryear{{Miller} et~al.,}{{Miller}
  et~al.}{2019}]{Miller2019}
{Miller} J.~M.,  et~al., 2019, \mn@doi [\prd] {10.1103/PhysRevD.100.023008},
  \href {https://ui.adsabs.harvard.edu/abs/2019PhRvD.100b3008M} {100, 023008}

\bibitem[\protect\citeauthoryear{{Miller}, {Sprouse}, {Fryer}, {Ryan},
  {Dolence}, {Mumpower}  \& {Surman}}{{Miller} et~al.}{2020}]{Miller2020}
{Miller} J.~M.,  {Sprouse} T.~M.,  {Fryer} C.~L.,  {Ryan} B.~R.,  {Dolence}
  J.~C.,  {Mumpower} M.~R.,   {Surman} R.,  2020, \mn@doi [\apj]
  {10.3847/1538-4357/abb4e3}, \href
  {https://ui.adsabs.harvard.edu/abs/2020ApJ...902...66M} {902, 66}

\bibitem[\protect\citeauthoryear{{M{\"o}sta}, {Roberts}, {Halevi}, {Ott},
  {Lippuner}, {Haas}  \& {Schnetter}}{{M{\"o}sta} et~al.}{2018}]{Mosta+17}
{M{\"o}sta} P.,  {Roberts} L.~F.,  {Halevi} G.,  {Ott} C.~D.,  {Lippuner} J.,
  {Haas} R.,   {Schnetter} E.,  2018, \mn@doi [\apj]
  {10.3847/1538-4357/aad6ec}, \href
  {https://ui.adsabs.harvard.edu/abs/2018ApJ...864..171M} {864, 171}

\bibitem[\protect\citeauthoryear{{Murray}}{{Murray}}{2011}]{murray2011}
{Murray} N.,  2011, \mn@doi [\apj] {10.1088/0004-637X/729/2/133}, \href
  {https://ui.adsabs.harvard.edu/abs/2011ApJ...729..133M} {729, 133}

\bibitem[\protect\citeauthoryear{{Nishimura}, {Kotake}, {Hashimoto}, {Yamada},
  {Nishimura}, {Fujimoto}  \& {Sato}}{{Nishimura} et~al.}{2006}]{Nishimura2006}
{Nishimura} S.,  {Kotake} K.,  {Hashimoto} M.-a.,  {Yamada} S.,  {Nishimura}
  N.,  {Fujimoto} S.,   {Sato} K.,  2006, \mn@doi [\apj] {10.1086/500786},
  \href {https://ui.adsabs.harvard.edu/abs/2006ApJ...642..410N} {642, 410}

\bibitem[\protect\citeauthoryear{Ott}{Ott}{2016}]{Ott2016}
Ott U.,  2016, Isotope Variations in the Solar System: Supernova Fingerprints.
Springer International Publishing, Cham, pp 1--27,
  \mn@doi{10.1007/978-3-319-20794-0_17-1}, \url
  {https://doi.org/10.1007/978-3-319-20794-0_17-1}

\bibitem[\protect\citeauthoryear{{Qian} \& {Woosley}}{{Qian} \&
  {Woosley}}{1996}]{Qian1996}
{Qian} Y.~Z.,  {Woosley} S.~E.,  1996, \mn@doi [\apj] {10.1086/177973}, \href
  {https://ui.adsabs.harvard.edu/abs/1996ApJ...471..331Q} {471, 331}

\bibitem[\protect\citeauthoryear{{Radice}, {Galeazzi}, {Lippuner}, {Roberts},
  {Ott}  \& {Rezzolla}}{{Radice} et~al.}{2016}]{Radice2016}
{Radice} D.,  {Galeazzi} F.,  {Lippuner} J.,  {Roberts} L.~F.,  {Ott} C.~D.,
  {Rezzolla} L.,  2016, \mn@doi [\mnras] {10.1093/mnras/stw1227}, \href
  {https://ui.adsabs.harvard.edu/abs/2016MNRAS.460.3255R} {460, 3255}

\bibitem[\protect\citeauthoryear{{Rosswog}}{{Rosswog}}{2005}]{Rosswog2005}
{Rosswog} S.,  2005, \mn@doi [\apj] {10.1086/497062}, \href
  {https://ui.adsabs.harvard.edu/abs/2005ApJ...634.1202R} {634, 1202}

\bibitem[\protect\citeauthoryear{{Sch{\"o}nrich} \& {McMillan}}{{Sch{\"o}nrich}
  \& {McMillan}}{2017}]{Schonrich17}
{Sch{\"o}nrich} R.,  {McMillan} P.~J.,  2017, \mn@doi [\mnras]
  {10.1093/mnras/stx093}, \href
  {https://ui.adsabs.harvard.edu/abs/2017MNRAS.467.1154S} {467, 1154}

\bibitem[\protect\citeauthoryear{Shibata \& Hotokezaka}{Shibata \&
  Hotokezaka}{2019}]{Shibata:2019wef}
Shibata M.,  Hotokezaka K.,  2019, \mn@doi [Ann. Rev. Nucl. Part. Sci.]
  {10.1146/annurev-nucl-101918-023625}, 69, 41

\bibitem[\protect\citeauthoryear{{Siegel}, {Barnes}  \& {Metzger}}{{Siegel}
  et~al.}{2019}]{Siegel+18}
{Siegel} D.~M.,  {Barnes} J.,   {Metzger} B.~D.,  2019, \mn@doi [\nat]
  {10.1038/s41586-019-1136-0}, \href
  {https://ui.adsabs.harvard.edu/abs/2019Natur.569..241S} {569, 241}

\bibitem[\protect\citeauthoryear{{Sneden}, {Cowan}  \& {Gallino}}{{Sneden}
  et~al.}{2008}]{Sneden2008}
{Sneden} C.,  {Cowan} J.~J.,   {Gallino} R.,  2008, \mn@doi [\araa]
  {10.1146/annurev.astro.46.060407.145207}, \href
  {https://ui.adsabs.harvard.edu/abs/2008ARA&A..46..241S} {46, 241}

\bibitem[\protect\citeauthoryear{{Takahashi}, {Witti}  \& {Janka}}{{Takahashi}
  et~al.}{1994}]{Takahashi1994}
{Takahashi} K.,  {Witti} J.,   {Janka} H.~T.,  1994, \aap, \href
  {https://ui.adsabs.harvard.edu/abs/1994A&A...286..857T} {286, 857}

\bibitem[\protect\citeauthoryear{{Tanaka} et~al.,}{{Tanaka}
  et~al.}{2017}]{Tanaka+2017}
{Tanaka} M.,  et~al., 2017, \mn@doi [\pasj] {10.1093/pasj/psx121}, \href
  {https://ui.adsabs.harvard.edu/abs/2017PASJ...69..102T} {69, 102}

\bibitem[\protect\citeauthoryear{{Tang}, {Liu}, {McKeegan}, {Tissot}  \&
  {Dauphas}}{{Tang} et~al.}{2017}]{Tang+2017}
{Tang} H.,  {Liu} M.-C.,  {McKeegan} K.~D.,  {Tissot} F. L.~H.,   {Dauphas} N.,
   2017, \mn@doi [\gca] {10.1016/j.gca.2017.03.001}, \href
  {https://ui.adsabs.harvard.edu/abs/2017GeCoA.207....1T} {207, 1}

\bibitem[\protect\citeauthoryear{{Vincent}, {Foucart}, {Duez}, {Haas},
  {Kidder}, {Pfeiffer}  \& {Scheel}}{{Vincent} et~al.}{2020}]{Vincent2020}
{Vincent} T.,  {Foucart} F.,  {Duez} M.~D.,  {Haas} R.,  {Kidder} L.~E.,
  {Pfeiffer} H.~P.,   {Scheel} M.~A.,  2020, \mn@doi [\prd]
  {10.1103/PhysRevD.101.044053}, \href
  {https://ui.adsabs.harvard.edu/abs/2020PhRvD.101d4053V} {101, 044053}

\bibitem[\protect\citeauthoryear{Wallner et~al.,}{Wallner
  et~al.}{2015}]{Wallner_2015}
Wallner A.,  et~al., 2015, \mn@doi [Nature Communications]
  {10.1038/ncomms6956}, 6

\bibitem[\protect\citeauthoryear{{Wallner} et~al.,}{{Wallner}
  et~al.}{2021}]{Wallner:2021}
{Wallner} A.,  et~al., 2021, \mn@doi [Science] {10.1126/science.aax3972}, \href
  {https://ui.adsabs.harvard.edu/abs/2021Sci...372..742W} {372, 742}

\bibitem[\protect\citeauthoryear{Wanajo, Sekiguchi, Nishimura, Kiuchi, Kyutoku
  \& Shibata}{Wanajo et~al.}{2014}]{Wanajo:2014wha}
Wanajo S.,  Sekiguchi Y.,  Nishimura N.,  Kiuchi K.,  Kyutoku K.,   Shibata M.,
   2014, \mn@doi [\apjl] {10.1088/2041-8205/789/2/L39}, 789, L39

\bibitem[\protect\citeauthoryear{{Wasserburg}, {Busso}, {Gallino}  \&
  {Nollett}}{{Wasserburg} et~al.}{2006}]{wasserburg2006}
{Wasserburg} G.~J.,  {Busso} M.,  {Gallino} R.,   {Nollett} K.~M.,  2006,
  \mn@doi [\nphysa] {10.1016/j.nuclphysa.2005.07.015}, \href
  {https://ui.adsabs.harvard.edu/abs/2006NuPhA.777....5W} {777, 5}

\bibitem[\protect\citeauthoryear{{Winteler}, {K{\"a}ppeli}, {Perego},
  {Arcones}, {Vasset}, {Nishimura}, {Liebend{\"o}rfer}  \&
  {Thielemann}}{{Winteler} et~al.}{2012}]{Winteler+12}
{Winteler} C.,  {K{\"a}ppeli} R.,  {Perego} A.,  {Arcones} A.,  {Vasset} N.,
  {Nishimura} N.,  {Liebend{\"o}rfer} M.,   {Thielemann} F.-K.,  2012, \mn@doi
  [\apjl] {10.1088/2041-8205/750/1/L22}, \href
  {http://adsabs.harvard.edu/abs/2012ApJ...750L..22W} {750, L22}

\bibitem[\protect\citeauthoryear{{Woosley}, {Wilson}, {Mathews}, {Hoffman}  \&
  {Meyer}}{{Woosley} et~al.}{1994}]{Woosley1994}
{Woosley} S.~E.,  {Wilson} J.~R.,  {Mathews} G.~J.,  {Hoffman} R.~D.,   {Meyer}
  B.~S.,  1994, \mn@doi [\apj] {10.1086/174638}, \href
  {https://ui.adsabs.harvard.edu/abs/1994ApJ...433..229W} {433, 229}

\bibitem[\protect\citeauthoryear{Wu, Fernández, Martínez-Pinedo  \&
  Metzger}{Wu et~al.}{2016}]{Wu:2016pnw}
Wu M.-R.,  Fernández R.,  Martínez-Pinedo G.,   Metzger B.~D.,  2016, \mn@doi
  [\mnras] {10.1093/mnras/stw2156}, 463, 2323

\makeatother
\end{thebibliography}

\appendix
\section{Milky Way Model}
\label{appA}
The model for the Milky Way is adapted from the best fit model from \citet{cote2019}. The star formation rate is assumed to be directly proportional to the gas mass with a star formation efficiency  $f_\star=2.6\times10^{-10}~{\rm yr}^{-1}$. The two-infall prescription of \citet{chiappini1997} is adopted for the gas inflow rate given by 
\begin{equation}
    \dot{M}_{\rm inflow}=A_1\exp\left(\frac{-t}{\tau_1}\right)+A_2\exp\left(\frac{t_{\rm max}-t}{\tau_2}\right),
\end{equation}
where $A_1=46\,M_\odot~{\rm yr}^{-1}$, $A_2=5.9\,M_\odot~{\rm yr}^{-1}$, $\tau_1=0.68$ Gyr, $\tau_2=7$ Gyr, and $t_{\rm max}=1$ Gyr. An initial time offset of 300 Myr is adopted to account for the lack of star formation in the very early Galaxy. The outflow rate is assumed to be proportional to the star formation rate with a proportionality constant $\eta=0.52$. The nucleosynthetic yields for low- and intermediate-mass stars are taken from \citet{karakas2010}. The massive star yields are taken from \citet{kobayashi2006} with half the stars in the range of $20$--$40~M_\odot$ assumed to explode as hypernovae. The number of SNe Ia per stellar mass of star formation is taken to be $10^{-3}$ with a DTD $\propto t^{-1}$ and a minimum delay time of $\sim 40$ Myr corresponding to the lifetime of a $8\,M_\odot$ star. A fraction $f_r=8\times10^{-4}$ of massive stars are assumed to form $r$-process events which corresponds to a current Galactic rate at $t_{\rm gal}=13.7$ Gyr of $\nu_0\sim10\,{\rm Myr}^{-1}$ and $\nu\sim15\,{\rm Myr}^{-1}$ at $t_\odot=9.2$ Gyr. 

%If you want to present additional material which would interrupt the flow of the main paper, it can be placed in an Appendix which appears after the list of references.

%%%%%%%%%%%%%%%%%%%%%%%%%%%%%%%%%%%%%%%%%%%%%%%%%%

% Don't change these lines
\bsp	% typesetting comment
\label{lastpage}
\end{document}